\documentclass{pinchcr}

\usepackage{cite}
\usepackage{bm}
\usepackage{epsfig}

\newcommand{\be}{\begin{equation}}
\newcommand{\ee}{\end{equation}}
\newcommand{\ba}{\begin{eqnarray}}
\newcommand{\ea}{\end{eqnarray}}

\newcommand{\bq}{{\bf q}}
\newcommand{\bx}{{\bf x}}
\newcommand{\by}{{\bf y}}
\newcommand{\bk}{{\bf k}}
\newcommand{\bz}{{\bf z}}

\title{Overview of different characterisations of dynamic heterogeneity}

\author{L. Berthier}
\affiliation{Laboratoire des Collo{\"\i}des, Verres
 et Nanomat{\'e}riaux, Universit{\'e} Montpellier 2 and 
CNRS UMR 5587,
34095 Montpellier, France}

\author{G. Biroli}
\affiliation{Institut de Physique Th{{\'e}o}rique, CEA, 
IPhT, 91191 Gif sur Yvette,
France and CNRS URA 2306}

\author{J.-P. Bouchaud}
\affiliation{Science \& Finance, Capital Fund Management
6-8 Bd Haussmann, 75009 Paris, France}

\author{R. L. Jack}
\affiliation{Department of Physics, 
University of Bath, Bath BA2 7AY, United Kingdom}

\begin{document}

\maketitle

\preface

Dynamic heterogeneity is now recognised as a central aspect 
of structural relaxation in disordered materials with slow dynamics, and 
was the focus of intense research in the last decade. Here 
we describe how initial, indirect observations of dynamic
heterogeneity have recently evolved into well-defined, 
quantitative, statistical characterisations, in particular through the use 
of high-order correlation and response functions. We highlight 
both recent progress and open questions about the characterisation 
of dynamic heterogeneity in glassy materials. We also discuss the limits
of available tools and describe a few candidates for future 
research in order to gain deeper understanding of the origin 
and nature of glassiness in disordered systems.

\maintext

\section{Introduction}

\subsection{Dynamical heterogeneity in glassy materials} 

The glass transition is often cited as a
profound outstanding problem in condensed matter physics.  
This field may be contrasted with that of simple liquids, for
which the broad picture is now well-established, and 
appropriate theoretical methods are well-developed~\protect\shortcite{hansen}. 
Why, then, is the glass problem so difficult?  

From a theoretical perspective,
a central difficulty arises from the importance of fluctuations
in glassy systems.  Both the liquid and the glass have disordered structure,
so even if all molecules in the system are identical, they experience
different local environments.  In the liquid, these differences
can be neglected: one may infer the behaviour of the system from
that of a typical particle in a typical environment.  Thus, for
example, microscopic properties, such as
the rate with which particles diffuse in the liquid, are directly related to
bulk properties, such as the viscosity.  However, as the glass
transition is approached, it becomes increasingly difficult to
characterise `typical' particles and `typical' environments
because a variety of different behaviors emerges. 
  Within a given interval of time,
some particles may move distances comparable to their size, while
others remain localised near their original positions.  Thus, on
these time scales, we can
refer to `mobile' and `immobile' particles. Of course, on long
enough time scales, ergodicity ensures that particles 
become statistically identical.

\begin{figure}[t]
\begin{center}
 \includegraphics[height=4.cm,clip]{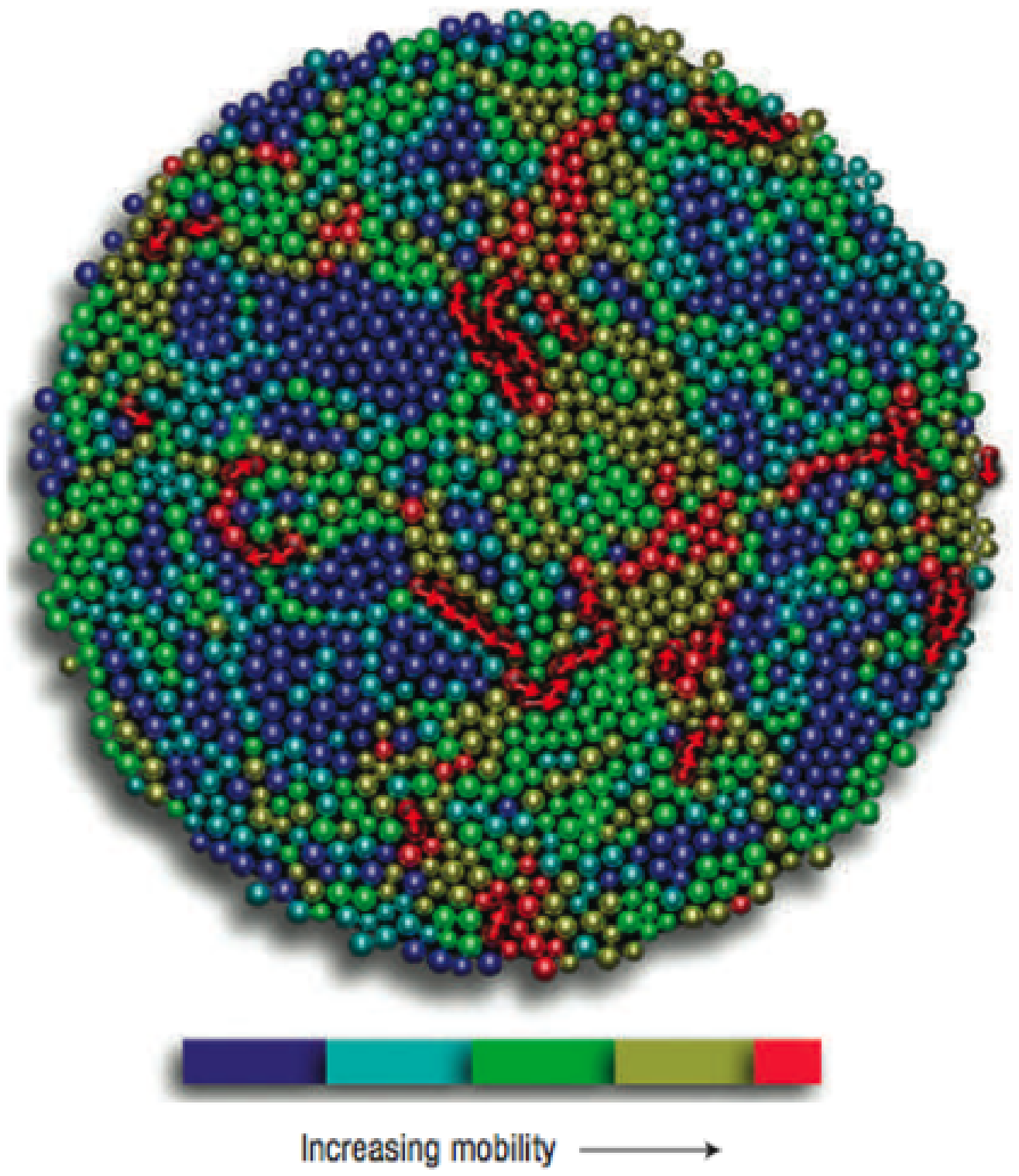}
 \includegraphics[height=3.7cm,clip]{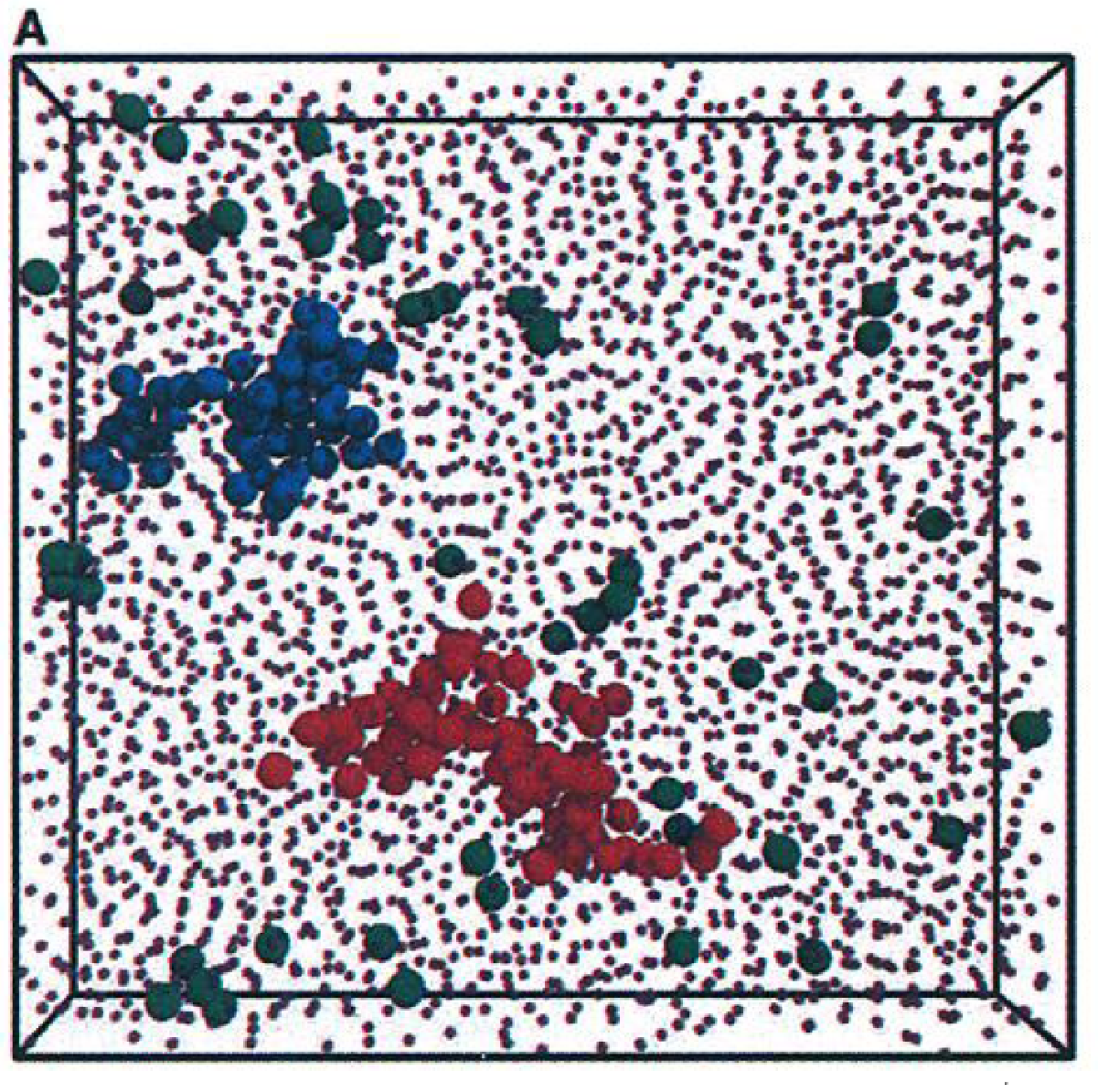}
 \includegraphics[height=3.6cm,clip]{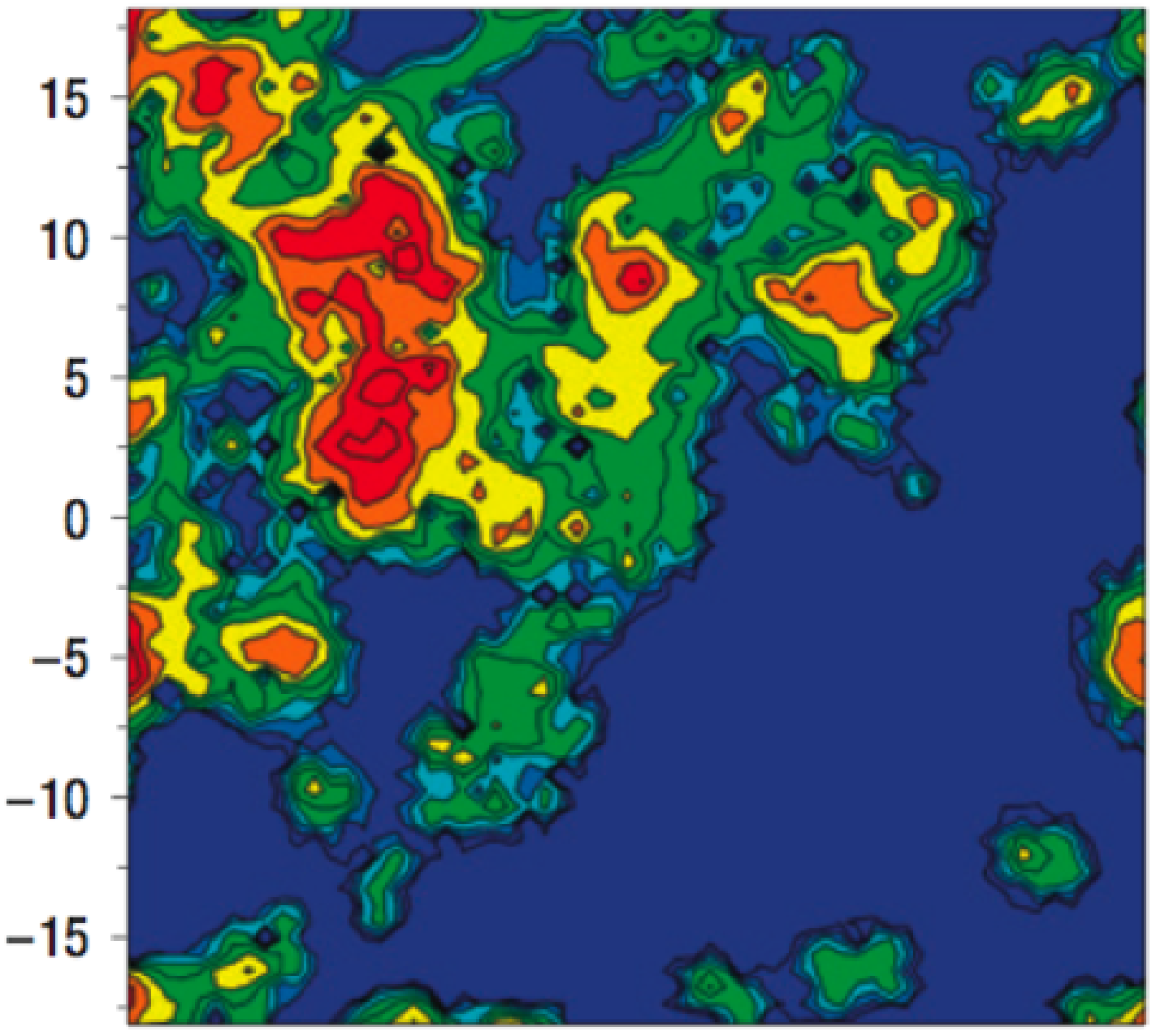}
\end{center}
\caption{Three examples of dynamical heterogeneity.  
In all cases, the figures highlight the clustering of
particles with similar mobility.
(Left)~Granular fluid of ball bearings, with a colour scale
showing a range of mobility increasing from 
blue to red~\protect\shortcite{durian}.  
(Centre)~Colloidal hard sphere suspension,
with most mobile particles highlighted~\protect\shortcite{Weeks}.  
(Right)~Computer simulation of a two-dimensional system of repulsive disks.
The colour scheme indicates the presence of particles for which
motion is reproducibly immobile or mobile, 
respectively from blue to red~\protect\shortcite{harrowell4}.} 
\label{fig:dh_poster}
\end{figure}

In case this discussion seems rather abstract, we refer the
reader to Fig.~\ref{fig:dh_poster}.  Here, we show several systems in which
mobile and immobile particles can be identified in particular
trajectories using different methods. Strikingly, these images reveal
that particles with different mobilities do not appear randomly in space
but are clustered. This observation suggests that structural 
relaxation in disordered systems is a nontrivial dynamical process.    
In its narrow sense, the term `dynamical heterogeneity' encapsulates 
the spatial correlations observed in Fig.~\ref{fig:dh_poster}.  
However, the term is frequently used in a broader sense,
referring to a range of fluctuation phenomena that arise
from deviations from the `typical' behaviour~\protect\shortcite{ediger}.

Over the last decade, it has become clear from experiments
and computer simulations that a variety of glassy
systems display the kind of clusters shown in 
Fig.~\ref{fig:dh_poster}.  Experimentally,
their existence can be inferred from experiments in 
molecular liquids~\protect\shortcite{ediger,reviewdh1,reviewdh2,reviewdh3}, 
while direct observation of single
particle motion makes them vivid in colloids 
and granular media~\protect\shortcite{Weeks,durian,kegel,dauchotbiroli}.
In computer simulations, spherical particles
with simple pair potentials have been used as models for both colloidal
and molecular systems, with dynamically heterogeneous
behaviour clearly present in a variety of models.  
Dynamical heterogeneity has also been investigated
in a large number of more schematic lattice models, such as 
kinetically constrained or lattice glass 
models~\protect\shortcite{Ritort-Sollich}.
This important body of experimental and computational 
observations has also stimulated important theoretical developments
since they represent a new set of observations against which 
existing theories can be confronted.   

So far, the reader may be unconvinced of the difficulty of the problem.
After all, theoretical methods for analysing systems with 
large fluctuations and correlated domains already exist: the methods
developed to describe critical phenomena, such as 
field theory and the renormalisation group.  
These ideas are indeed central to this chapter, but their 
application to glassy liquids has required substantial new 
insight into the nature of the relevant fluctuations
and observables.
The reason is that the distinction between mobile and immobile particles
is in essence dynamical. Therefore, if one analyses static snapshots of 
viscous liquids, there is little evidence of increasing 
fluctuations or heterogeneity, at least when analysed using
standard liquid state correlation functions.
Instead of an ensemble of snapshots, one must apply the methods of critical
phenomena to an ensemble of `movies' (i.e. trajectories, or dynamical 
histories) of the system. This will be the approach that we will follow
in later sections.  

\subsection{Suitable probes for
the emergence of glassiness}

Returning to Fig.~\ref{fig:dh_poster},
the observation of clusters of mobile particles (or at least 
of regions with correlated mobility) raises many questions.
What is the nature of the clusters?
Indeed, are they of the same nature in each case?  What
determines their size?  How is their size distribution related to the
relaxation time of the system, if at all?  Do the particles in fast or
slow clusters have different local environments that can be
characterised by any simple structural measure?  From a theoretical
point of view, it is natural to ask whether the correlated
regions represent the cooperatively rearranging regions 
imagined long ago by Adam-Gibbs~\protect\shortcite{AGpaper}; 
whether they mirror
some soft elastic modes in the system; how they might connect
to locally-ordered domains; or whether they reveal the presence
of localised defects that facilitate structural rearrangement.

However, before addressing these ambitious questions, we must answer a
more prosaic one.  How can the `clusters of mobile and 
immobile particles' be defined and 
characterised? In Fig.~\ref{fig:dh_poster}, 
the mobile particles were
identified by a threshold on their displacement,
over a particular time scale.  Indeed, 
many early papers have suggested different ways to define
`mobility' and `clusters' that are similar in spirit but 
different in practice, as we detail in the following sections. However,
one must certainly evaluate
how strongly the cluster properties depend on thresholds or
time scales, which might hinder firm conclusions and may
prevent fair comparison between different systems. 

In more recent years, 
it has become possible to define and measure observables that 
can be determined without arbitrariness in a range of systems, 
are amenable to analytic theory and scaling arguments, 
and may sometimes be inferred from experimental data
even in molecular liquids. These are known as `four-point
correlation functions' and are now broadly accepted as standard tools
for analysing dynamical heterogeneity.

Within this toolbox, a central role is played by the four-point
dynamical susceptibility $\chi_4(t)$.  Loosely speaking,
it measures the number of particles involved in correlated 
motion on times scales of the order of $t$.
To interpret the dynamical susceptibility, 
it is useful to invoke an analogy with
critical phenomena, in which a (static) correlation length $\xi$
diverges at a critical temperature $T_c$, accompanied by the spontaneous
appearance of an order parameter.  This divergence
is associated with a diverging susceptibility $\chi$, which
may be measured either through the fluctuations of the order
parameter or through the response of the order parameter to its
conjugate field.  
When considering such phenomena in glasses,
a problem arises, in that a static order parameter and its
conjugate field are not known. Instead, a fairly good dynamical order parameter
is given by any generic dynamical two-point correlator, e.g. 
density-density, displaying the slowing down
of the dynamics. This motivated the use 
of a four-point dynamical susceptibility associated to 
spontaneous fluctuations of the dynamical order parameter. 
This allows the identification of a dynamic length scale, $\xi_4(t)$, 
and a susceptibility, $\chi_4(t)$, by analogy
with conventional critical phenomena.  

Returning to Fig.~\ref{fig:dh_poster},
simulation studies and recent experiments indicate that
the clustering of mobile particles is directly linked with an increasing
susceptibility $\chi_4$ and an increasing 
correlation length scale $\xi_4$.  
The central part of this chapter will be devoted to a discussion of
these four-point functions. We will discuss some of the
insights that they have revealed into the nature of glassy behaviour in
liquids, colloids, and granular media.
However, these studies also revealed that interpretation of 
four-point functions may be somewhat ambiguous, while direct 
measurements of correlation 
length scales remain difficult. 
Additionally, the averaging procedure inherent in the four-point
functions means that they may obscure important features of 
dynamic heterogeneity such as
the cluster shape and the nature of interfaces between clusters of
mobile and immobile particles. Towards the end of the chapter, we will
discuss a range of alternative observables that complement the information
available from four-point functions.

\section{Observables for characterising dynamical heterogeneity}

\subsection{Two-point observables and their inadequacy}
\label{sec:inadequate}

We begin with a review of some 
two-point functions that are used to characterise simple liquids.
We will show that these functions are largely blind to the 
dynamically heterogeneous behaviour shown in Fig.~\ref{fig:dh_poster},
motivating the discussion of more discriminating 
observables\footnote{There 
are 
other examples of systems for which standard two-point correlators are
blind to interesting intermittent (or heterogeneous) effects, such as 
turbulence or 
financial markets, and for which higher order correlations are informative.}.

For any fluid of particles, a natural quantity to measure is the
structure factor,
\be
S(q)= \frac{1}{N}
\left\langle  \rho_{\bf q}(t) \rho_{\bf -q} (t)\right\rangle, 
\label{sofq}
\ee
where brackets indicate an ensemble average, and 
the Fourier component of the density is
\be
\rho_{\bf q} (t) = \sum_{i=1}^N e^{i {\bf q} \cdot {\bf r}_i(t)},
\ee
with $N$ being the number of particles 
and ${\bf r}_i(t)$ being the position of particle $i$ at time $t$. The structure
factor gives information about the strength of density fluctuations on
a length scale $2\pi / |{\bf q}|$.  However, its behaviour in the vicinity
of the glass transition is unremarkable, with no hint of the dynamic  
clustering of Fig.~\ref{fig:dh_poster}. Although static heterogeneities
in the density would directly imply  the existence of dynamic
heterogeneity, the reverse is not true. 
Thus, dynamic heterogeneity related to the motion of particles has 
a much more subtle origin.
Note that more complicated static correlation functions have been 
studied~\protect\shortcite{debenedetti}, especially in numerical work, 
all attempting to identify fluctuations of some prescribed sort 
of local order (translational, orientational, etc.). 
Until now, there are no strong indications of a diverging, 
or even 
substantially growing, lengthscale~\protect\shortcite{static1,static2}. 
In order to see the growth of a static amorphous correlation, one
possibility is
to introduce the so-called ``point-to-set'' correlation function that
we discuss later, see section \ref{pointtoset}.

\begin{figure}[t]
\begin{center}
 \psfig{file=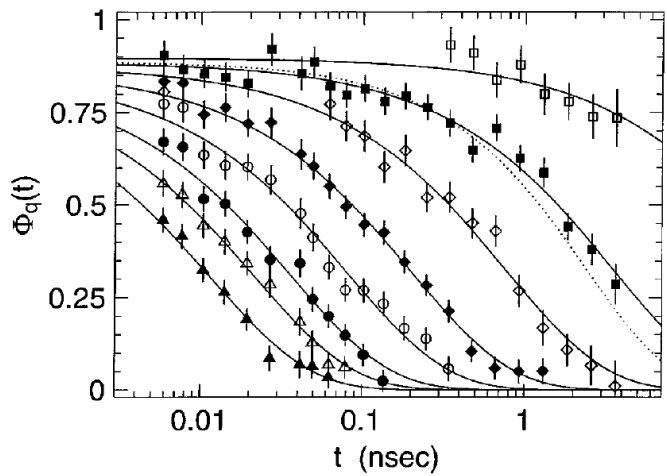,width=9.cm}
\end{center}
\caption{\label{fqt} Temperature evolution of the normalised 
intermediate scattering
  function, $\phi_{\bf q}(t) = S({\bf q},t)/S({\bf q},0)$,
for supercooled
  glycerol~\protect\shortcite{glycerol}. 
  Temperatures decrease from 413~K to 270~K from left to right.
  The lines are fits with a stretched exponential form.}
\end{figure}

We therefore turn to dynamical observables.
A quantity relevant for light and neutron scattering 
experiments is the
intermediate scattering function, 
\begin{equation}
F({\bf q},t) = \frac 1 N
\left\langle  \rho_{\bf q}(t) 
\rho_{\bf -q}(0)  
\right\rangle.
\label{isf}
\end{equation}
Measurements of this function
by neutron scattering in supercooled
glycerol~\protect\shortcite{glycerol} 
are shown for different temperatures in Fig.~\ref{fqt}.
These curves suggest a first, rather fast, relaxation to a plateau followed
by a second, much slower, relaxation. 
The plateau is due to the fraction of density fluctuations that 
are frozen on intermediate timescales, but
eventually relax during the second relaxation. The latter is called 
`alpha-relaxation', and corresponds to the structural relaxation 
of the liquid. The plateau is akin to the Edwards-Anderson order parameter, 
defined for spin glasses which measures the fraction 
of frozen spin fluctuations~\protect\shortcite{binderkob}. 
Note that the Edwards-Anderson parameter continuously increases from zero 
below the critical temperature in the conventional 
spin glass transition~\protect\shortcite{zebeyond}, while 
for structural glasses, a finite plateau value seems to 
emerge above any putative transition.

The full decay of the 
intermediate scattering function can be measured only within
a relatively small range of temperatures. 
In order to track the dynamic slowing down from microscopic 
to macroscopic timescales, other correlators have been studied.
A very popular one is the measurement of the dielectric linear
susceptibility which can be followed over 
up to 18 decades of frequency~\protect\shortcite{lunkenheimer}.  
It is generally accepted 
that different dynamic probes 
reveal similar temperature dependences for the relaxation time, 
at least as long as the probes measure local motion. 
In broad terms, the essential features on supercooling are a dramatic
increase in the correlation time, and a broad distribution of
time scales in the system characterised in the time-domain by 
non-exponential relaxation functions. 

It is increasingly accepted that the presence
of a such broad distributions of time scales in glassy systems  is
associated with the presence of mobile and immobile domains.
However, the size and shape of these domains, or even their 
very existence, cannot be deduced directly from $F({\bf q},t)$.
We are therefore motivated to consider more advanced
correlation functions.

\subsection{Indirect evidence: Intermittency and decoupling phenomena}

\begin{figure}[t]
\begin{center}
 \psfig{file=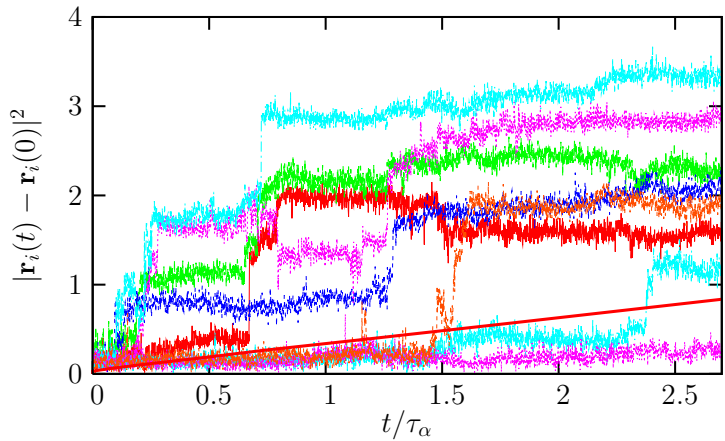,width=9.5cm}
\end{center}
\caption{\label{msd2} Time resolved squared displacements of individual 
particles in a simple model of a glass-forming liquid composed 
of Lennard-Jones particles~\protect\shortcite{lj-mc}. 
The average is shown as a smooth full line
and time is expressed in units of structural relaxation
time $\tau_\alpha$.  
Trajectories are composed of long periods of time during which particles
vibrate around well-defined positions, separated by rapid jumps that
are statistically widely distributed in time, underlying the importance of 
dynamic fluctuations.}
\end{figure}

We now return to the presence of mobile and immobile particles
in the supercooled phase.  
Simulation studies are ideal for studying fluctuation properties,
since accurate trajectories for all particles are accessible,
over long time scales.
These features are displayed in Fig.~\ref{msd2}, which shows that 
while the averaged
mean-squared displacements are smooth
functions of time, time signals for individual
particles clearly exhibit specific features that 
are not observed unless dynamics is resolved both in space and time. 
In this figure,
we observe that particle trajectories are very intermittent, being 
composed of a succession of long periods of time 
where particles simply vibrate around 
well-defined locations, separated by rapid `jumps'.  
Vibrations were previously inferred from the plateaux observed 
at intermediate times in the mean squared displacements or
intermediate scattering functions,
but the existence of jumps that are 
statistically widely distributed in time cannot be revealed 
from averaged quantities only.  
The fluctuations in Fig.~\ref{msd2} suggest,
and direct measurements confirm, the importance played by 
fluctuations around the averaged dynamical behaviour
to understand structural relaxation in glassy materials.

Remaining at the single particle level, these fluctuations
can be characterised through the (time-dependent) distribution 
of particle displacements.  This is the self-part
of the van-Hove function, defined as
\begin{equation}
G_s({\bf r},t) = \left\langle 
\frac{1}{N} \sum_{i=1}^N \delta ({\bf r} - [{\bf r}_i(t) - 
{\bf r}_i(0)] ) \right\rangle . 
\end{equation}
For an isotropic Gaussian diffusive process, one has 
$G_s({\bf r},t) \sim
\exp\left(-\frac{|{\bf r}|^2}{4 D_s t}\right)$.
While simple liquids are well-described by
such a distribution, simulations of glassy systems instead reveal strong 
deviations from Gaussian behaviour on the timescales 
relevant for structural relaxation~\protect\shortcite{glotzerkob}. 
In particular they reveal 
`broad' tails in the distributions that are much wider than expected 
from the Gaussian approximation. These tails are in fact 
well described by an exponential, rather than 
Gaussian, decay in a wide time window comprising the 
structural relaxation, 
\be
G_s({\bf r},t) \sim \exp \left( - \frac{|{\bf r}|}{\lambda(t)} \right),
\ee
which is a direct consequence of the intermittent 
motion shown in Fig.~\ref{msd2}~\protect\shortcite{pinaki}.
These tails reflect the existence of a 
population of particles that moves distinctively further 
than the rest and appears therefore to be much more
mobile.  This observation implies that relaxation 
in a viscous liquid differs qualitatively from that of a normal liquid 
where diffusion is close to Gaussian, 
and that a non-trivial statistics of single
particle displacements exists in materials with glassy 
dynamics.

\begin{figure}[t]
\begin{center}
 \psfig{file=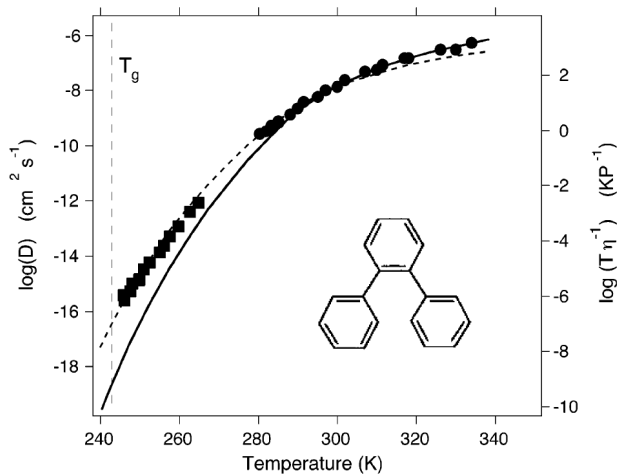,width=8.5cm}
\end{center}
\caption{\label{otp} Decoupling between viscosity (full line) and 
self-diffusion coefficient (symbols) in supercooled 
ortho-terphenyl~\protect\shortcite{edigerotp}.
The dashed line shows a fit with a `fractional' Stokes-Einstein 
relation, $D_s \sim (T / \eta)^\zeta$ with $\zeta \sim 0.82$ 
instead of the expected value $\zeta=1$.}
\end{figure}

Another influential phenomenon that was related early on 
to the existence of dynamic heterogeneity
is the decoupling of self-diffusion ($D_s$) and 
viscosity ($\eta$). In the high temperature
liquid, self-diffusion and viscosity are related by the 
Stokes-Einstein relation~\protect\shortcite{hansen}, 
$D_s \eta / T = const$. 
For a large particle moving in a fluid the constant is equal to $1/(6\pi R)$
where $R$ is the particle radius. Physically, the Stokes-Einstein relation 
means that two different measures of the relaxation time, $R^2/D_s$ and $\eta
R^3/T$, lead to the same timescale up to a constant factor. In supercooled
liquids this phenomenological law 
breaks down, as shown in Fig.~\ref{otp} for 
ortho-terphenyl~\protect\shortcite{edigerotp}. It is commonly found that $D_s^{-1}$ 
does not increase as fast as $\eta$ so that,  
at $T_g$, the product $D_s \eta$ has significantly increased as 
compared to its Stokes-Einstein value.
The Stokes-Einstein `violation' factor is larger for more fragile 
liquids, and can be as high as $10^3$.
This phenomenon, although less spectacular than the overall change of 
viscosity, is a strong indication that different ways to measure 
relaxation times
lead to different answers and, thus, is a strong hint of the existence
of broad distributions of relaxation 
timescales~\protect\shortcite{stillinger,Gilles}.

Indeed, a natural explanation of this effect is that different observables 
probe the underlying distribution of relaxation 
times in different ways~\protect\shortcite{ediger}. 
For example, the self-diffusion coefficient of tracer particles is dominated 
by the more mobile particles whereas the viscosity or other measures of
structural relaxation probe the timescale needed for every particle
to move. An unrealistic but instructive example is a model where
there is a small, non-percolative subset of particles that are 
blocked forever, coexisting with a majority of mobile 
particles. In this case, the structure never fully relaxes but the
self-diffusion coefficient is non-zero because of the mobile particles. 
Although unrealistic since all particles move in a viscous liquid, 
this example shows how different observables are likely to probe 
different moments of the distribution of
timescales, as explicitly shown within several theoretical 
frameworks~\protect\shortcite{Gilles,jung,maibaum,heuer08}. 

\subsection{Early studies of dynamic heterogeneity}

The phenomena described above, although certainly an indication 
of spatio-temporal fluctuations, do not allow one to study
how these fluctuations are correlated in space.  
However, this is a fundamental issue both from the experimental and
theoretical points of view, as discussed in the introduction.
To discriminate between different explanations of glassy behaviour, 
it would be useful to know:
How large are the regions that are faster or
slower than the average? How does their size depend on temperature? Are these
regions compact or fractal?    

These important questions were first addressed in pioneering works using 
four-dimensional NMR~\protect\shortcite{nmr,nmr2}, 
and by directly 
probing fluctuations at the nanoscopic scale using microscopy techniques. 
In particular, Vidal Russel and
Israeloff used Atomic Force Microscopy techniques \protect\shortcite{israeloff} to measure 
the polarisation fluctuations in a volume of size of few tens of nanometers 
in a supercooled polymeric liquid (PVAc) close to $T_g$. 
In this spatially resolved measurement, the hope is to probe a small enough
number of dynamically correlated regions, and to
detect their dynamics. 
Indeed, the time signals shown in Ref.~\protect\shortcite{israeloff} show a 
very intermittent dynamics, switching between
moments with intense activity, and moments with no dynamics 
at all, suggesting that extended regions of space 
indeed transiently behave as fast and slow regions.   
A much smoother signal would have been measured 
if dynamically correlated `domains' 
were not present.

Spatially resolved and NMR experiments are quite difficult.
They give undisputed information about the typical 
lifetime of the dynamic heterogeneity, but their determination of 
a dynamic correlation lengthscale is 
rather indirect, and has been performed on a small number of 
liquids in narrow temperature windows.
Nevertheless, an agreed consensus is that these experiments
reveal the existence of a non-trivial dynamic correlation length 
emerging at the glass transition, where it reaches a value of the 
order of 5 to 10 molecule diameters~\protect\shortcite{ediger}.      

\begin{figure}[t]
 \includegraphics[height=4.3cm,clip]{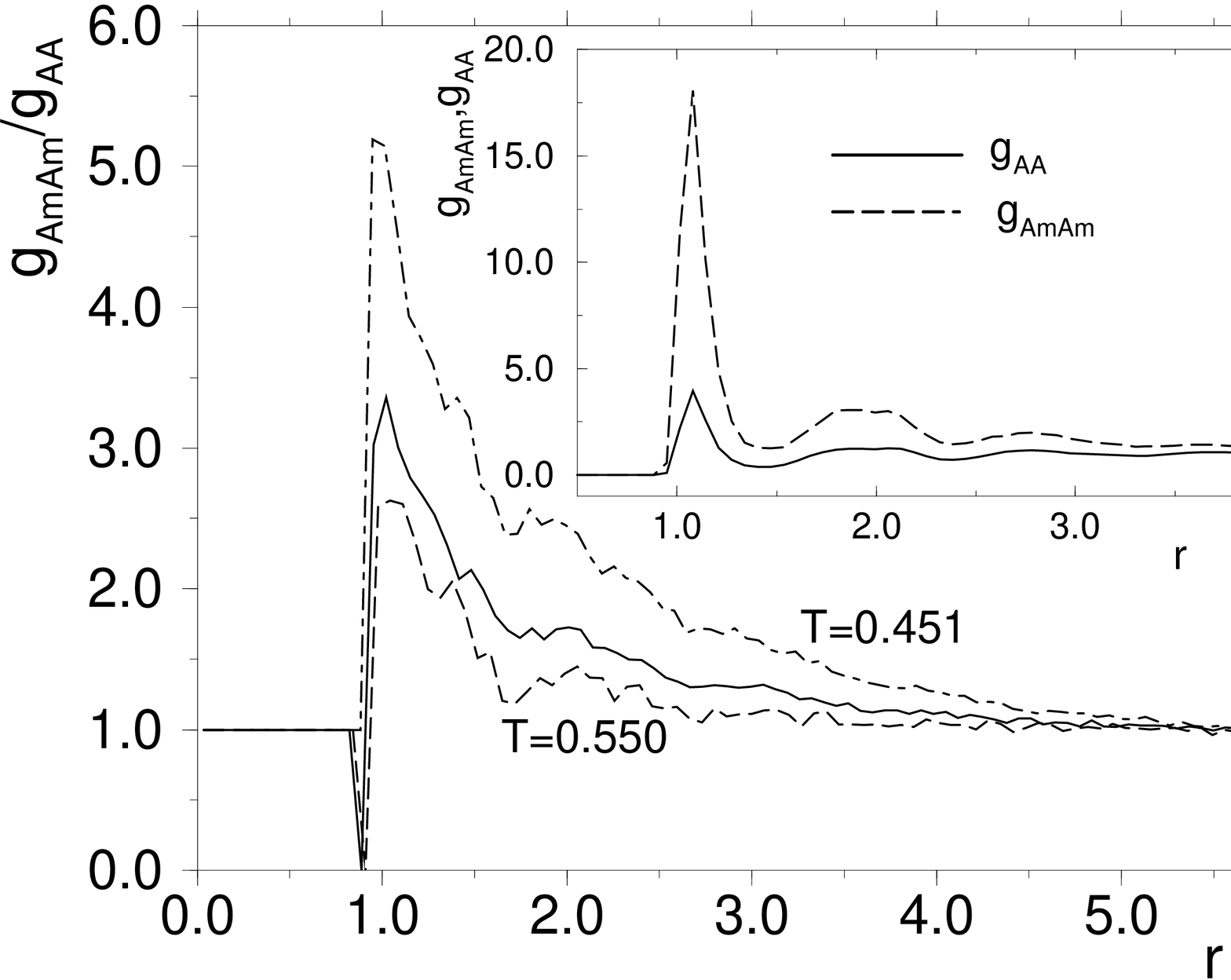}
\hspace*{-.2cm}
 \includegraphics[height=4.4cm,clip]{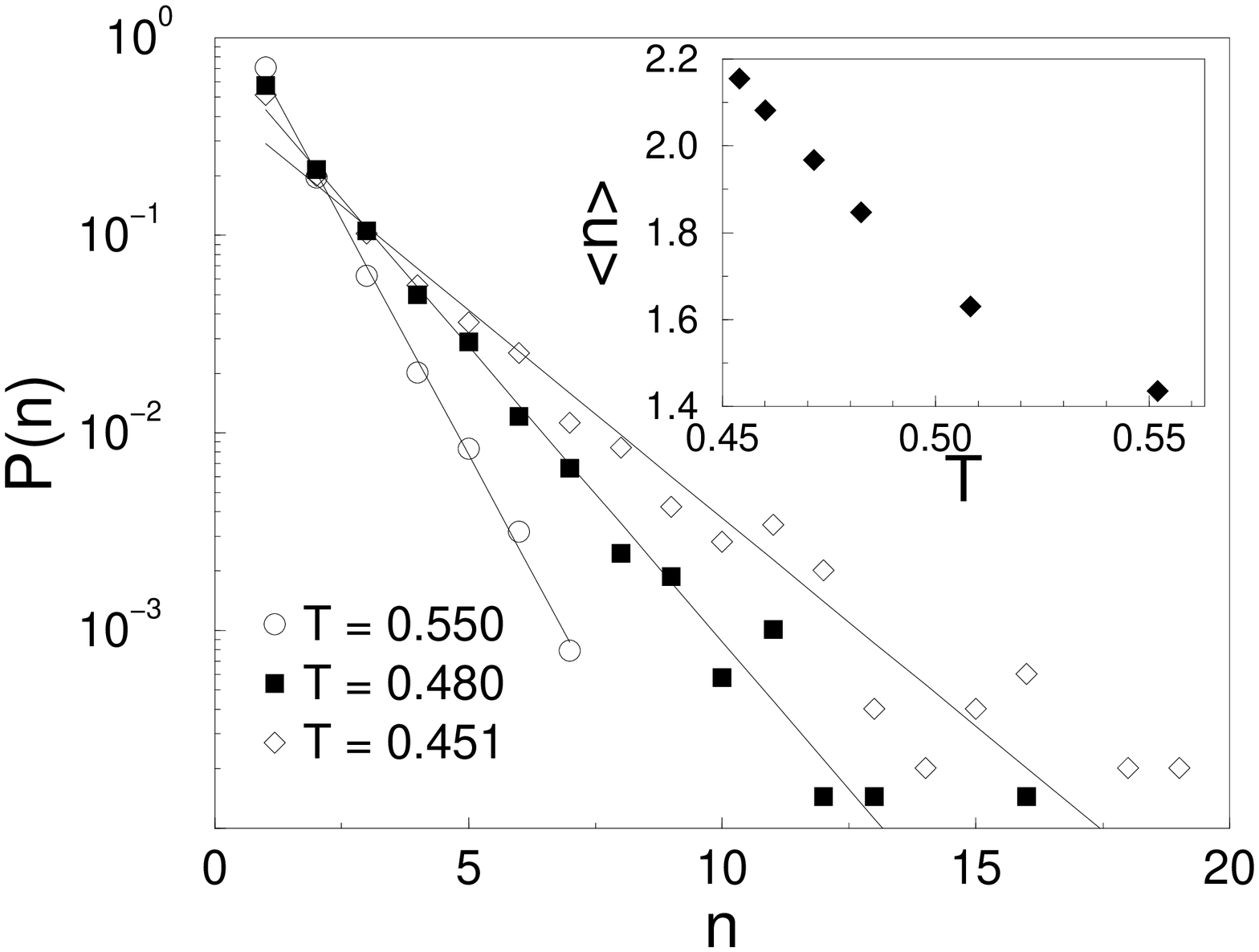}
\caption{Early measurements of dynamical heterogeneity
from computer simulation.  (Left)~The main figure
shows the probability that a particle is within
the active population, given that it is a distance
$r$ from another active particle~\protect\shortcite{glotzerkob}.  The increasing
correlation between active particles is apparent
on decreasing the temperature.  (Right)~The length distribution
of string-like clusters of active particles~\protect\shortcite{Donati98}.
}\label{fig:threshold_etc}
\end{figure}

More recently, studies of systems for which single-particle
dynamical data is available have led to direct observation
of clusters of mobile particles, such as those shown in 
Fig.~\ref{fig:dh_poster}. This is possible in materials
where the particles are big enough to be visualised
through microscopy, such as colloids, or with a camera, such 
as granular materials. 
In early studies, mobile and immobile particles
were usually identified
by using a threshold on their displacement within a given time
interval, dividing the particles into sub-populations.  This then allows
characterisation of correlations within the populations,
as shown in Fig.~\ref{fig:threshold_etc}.  For example,
ratios of radial distribution functions give the probability
that particles in the vicinity of a mobile particle are
themselves mobile. Other work concentrated for instance 
on the morphology and size of clusters of mobile particles, 
as shown in Fig.~\ref{fig:threshold_etc}.

Results such as those of Fig.~\ref{fig:threshold_etc} gave clear evidence
of large fluctuations and dynamical heterogeneity. 
However, unambiguous identification of a mobile population of particles 
proved difficult, with most distributions of mobility showing
broad but unimodal distributions. Similarly, the identification
of connected clusters of mobile particles introduces further ambiguity into the
data processing, with their size and shape depending quite strongly
on the definition of the mobile population.  To achieve unambiguous
and system-independent definitions of length scales and correlation
volumes, it is useful to define correlation functions that
do not involve separating particles into distinct populations,
nor the identification of connected clusters.  Four-point functions
are a natural choice in this regard, and will be discussed in the next
chapter.  

\subsection{Higher order correlations: four-point functions}

\subsubsection{Definitions}

In the previous section, we considered the probability that a
particle is within some mobile population, given that it has a mobile
particle a distance $r$ away (recall Fig.~\ref{fig:threshold_etc}).  
While the measurement of such a probability requires
the identification of a mobile population, there is a straightforward
alternative which contains similar information.  We first define
a continuously varying `mobility' $c_i(t,0)$ which indicates how
far or how much particle $i$ moves between times $t=0$ and $t$.  
Then, given two particles at separation $r$, one can measure the
degree to which their mobilities are correlated.  To this end,
it is convenient to define a `mobility field' 
through~\protect\shortcite{glotzer,poole2,glotzer3}
\be
c({\bf r} ; t,0) = \sum_i c_i(t,0) \delta({\bf r}-{\bf r}_i).
\ee  
Then, the spatial
correlations of the mobility are naturally captured through
the correlation function~\protect\shortcite{dasgupta-chi4}
\be
G_4(r;t) = \langle c({\bf r};t,0) c({\bf 0};t,0) 
\rangle - \langle c({\bf r};t,0) \rangle^2,
\label{equ:g4def}
\ee
which depends only on the single time $t$ and the single distance
$r = |{\bf r}|$ as long as the
average is taken at equilibrium in a translationally invariant
system.
The analogy with fluctuations in critical systems 
becomes clear in Eq.~(\ref{equ:g4def}) if one considers
the mobility field $c({\bf r};t,0)$ as playing the role of the 
order parameter for the transition, characterised by 
non-trivial fluctuations and correlations near the glass transition.  

Often, the mobility $c_i(t,0)$ is itself a two-point function. 
For example, to measure mobility on a length
scale $2 \pi / q$, one might consider $o_i(q,t)=e^{i{\bf q}\cdot {\bf r}_i(t)}$
and $c_i(t,0)=o_i({\bf q},t) o_i(-{\bf q},0)$.  In this case,
$o_i({\bf q},t)$ is related to a Fourier component of the density
of the system, and the average of $c_i(t,0)$ is the self-part
of the intermediate scattering function $F({\bf q},t)$
defined in Eq.~(\ref{isf}). Moving from a particle
observable $o_i({\bf q},t)$ to a field $o({\bf r};{\bf q},t)$, one arrives at
\be
G_4(r;t) = \langle o({\bf r};{\bf q},t) o({\bf r};-{\bf q},0) 
o({\bf 0};{\bf q},t) o({\bf 0};-{\bf q},0) \rangle
 - \langle o({\bf r};{\bf q},t) o({\bf r};-{\bf q},0) \rangle^2.
\label{equ:o4}
\ee
This correlation function is quartic in the operator $o$, so it
is known as a `four-point function'.  It measures correlations
on a length scale $r$, associated with motion between time zero
and time $t$; it depends additionally on the length scale $q^{-1}$
used in the definition of the particle mobility $c_i(t,0)$.
Since structural relaxation typically involves particle motion 
over a distance comparable to the particle size $R$, 
one typically chooses $q \sim 1/R$ and studies the remaining
$t$ and $r$ dependences.  

This definition of a real-space correlation function of the mobility
represents a vital advance in the characterisation of dynamical
heterogeneity.  In particular, it allows the language of field
theory and critical phenomena to be used in studying
dynamical fluctuations in glassy systems.  By analogy with 
critical phenomena, if there is a single dominant length scale
$\xi_4$ then one expects that for  large-$r$, the correlation
function decays as 
\be 
G_4(r; t) \sim \frac{A(t)}{r^{p}} e^{-r/\xi_4(t)},
\ee
with $p$ an exponent whose value is discussed below.
It is also natural to define the susceptibility  associated with
the correlation function
\be
\chi_4(t) = \int\mathrm{d}r\, G_4(r;t).
\ee
If the prefactor $A(t)$ were known, the susceptibility $\chi_4(t)$
could be used to extract the typical number of particles
involved in correlated motion.  That is, $\chi_4(t)$ may
be interpreted as the size of the correlated
clusters in Fig.~\ref{fig:dh_poster}.

Further, $\chi_4(t)$
can also be obtained from the fluctuations of 
the total mobility $C(t,0)=\int\mathrm{d}^d{\bf r}\, c({\bf r};t,0)$, 
through
\be
\chi_4(t) = N [ \langle C(t,0)^2 \rangle - \langle C(t,0) \rangle^2 ].
\label{equ:chi4var}
\ee
In practice, this formula allows an efficient measure of 
the degree of dynamical heterogeneity, at least in 
computer simulations and in experiments where the dynamics 
can be spatially and temporally resolved. 
As long as the observable $c({\bf r};t,0)$ is chosen appropriately, this
observable can be measured in a wide variety of systems, and serves
as a basis for fair comparisons of the extent of dynamical heterogeneity.

\subsubsection{The spin glass perspective: 
four-point functions in space and time}
\label{spin glass perspective}

It is interesting to note that four-point functions have
their origin in spin glass physics, where they were
used to investigate the onset of long-ranged amorphous order.
The key insight is due to Edwards and Anderson~\protect\shortcite{EA}.  
In a spin glass, one considers a set of $N$ localised degrees
of freedom, the spins, 
which we denote by $s_x$, with $x$ denoting the position
in space.  The spins interact by quenched random couplings.
Then, two-point correlation functions between spins 
such as $\sum_x \langle s_x s_{x+r}\rangle$ typically vanish for $r\neq0$,
since sites separated by a distance
$r$ may be either correlated or anti-correlated, with equal probability.
The Edwards-Anderson solution is to consider 
instead the static spin-glass correlation
$\chi_{SG} = N^{-1} \sum_x \langle s_x s_{x+r}\rangle^2$
which receives a positive contribution for both correlated and anti-correlated
sites.  

If one then considers a four-point dynamic function such as
\be
G_4 (r;t) = \frac{1}{N} \sum_x \langle s_x(0) s_{x+r}(0) 
s_x(t) s_{x+r}(t) \rangle,
\ee
then this approaches the static spin-glass correlation at 
long times.  In spin glasses, spontaneous symmetry breaking occurs into
an amorphous solid with long-ranged order at a second order spin glass 
transition. Near such a phase transition, $G_4(r;t)$
obeys critical scaling behaviour similar to that shown by
static correlation functions near familiar phase
transitions, and the corresponding susceptibility $\chi_4(t)$
diverges. 

In structural glasses, the possibility of a static transition to an amorphous
ordered state remains highly controversial.  However, four-point
functions such as $G_4(r;t)$ and $\chi_4(t)$ can be usefully employed
in the fluid state to characterise dynamical fluctuations, regardless
of their possible connection to any critical point. 

\subsubsection{Four-point susceptibilities in molecular,
colloidal and granular glasses}

\begin{figure}[t]
\begin{center}
 \psfig{file=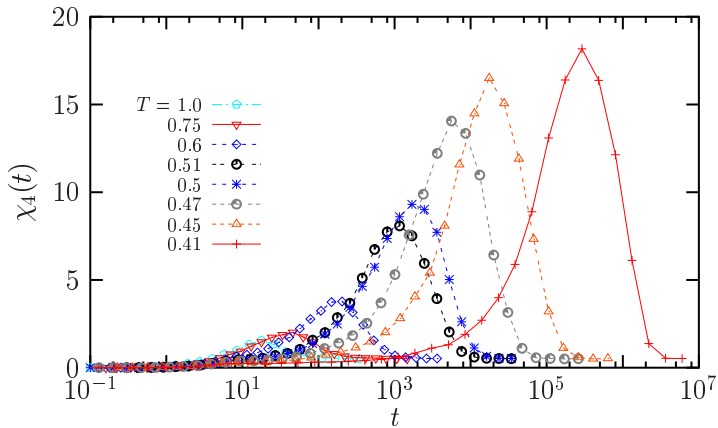,width=9.5cm}
\end{center}
\caption{\label{chi4ludo} Time dependence of $\chi_4(t)$ 
quantifying the spontaneous fluctuations of the 
self-intermediate scattering 
function in a molecular dynamics simulation of a 
Lennard-Jones supercooled liquid~\protect\shortcite{berthier}.
For each temperature, $\chi_4(t)$ has a maximum, which shifts to
larger times and has a larger value when $T$ is decreased, 
revealing the increasing lengthscale of dynamic heterogeneity in 
supercooled liquids approaching the glass transition.}
\end{figure}

The function $\chi_4(t)$ has been measured in computer simulations
of many different glass-forming liquids, by molecular dynamics,
Brownian and Monte Carlo 
simulations, see e.g.
\protect\shortcite{glotzer,poole2,glotzer3,glotzerfranzparisi,lacevic,berthier,glotzersilica,berthiersilica,parisi,castillo}. 
An example is shown in Fig.~\ref{chi4ludo} for a Lennard-Jones liquid.
The qualitative behaviour is similar in all cases
\protect\shortcite{franzparisi,TWBBB,jcpI}: 
as a function of time $\chi_4(t)$ first
increases, it has a peak on a timescale that tracks the structural relaxation
timescale and then it decreases. As mentioned above, 
the decrease of $\chi_4(t)$ at long times 
constitutes a  major difference with spin glasses. In a spin glass, 
$\chi_4$ would be a  monotonically increasing 
function of time whose long-time limit coincides with the static spin
  glass susceptibility. Physically, the difference is that 
 the spin glass state which 
  emerges at the transition is critical or marginally stable, i.e. 
  characterized by singular static responses.
  
The peak value of $\chi_4(t)$ measures the volume 
over which the structural relaxation processes are correlated.
Therefore, the most important result obtained from data
such as presented in Fig.~\ref{chi4ludo} is the temperature 
evolution of the peak height, which is found to increase when 
the temperature decreases and the dynamics slows down.
From such data, one gets direct evidence that the 
approach to the glass transition is accompanied by the 
development of increasingly long-ranged spatial
correlations of the dynamics.  
Note that if the dynamically correlated regions were compact, the peak 
of $\chi_4$ would be proportional to $\xi_4^d$ (in $d$ spatial dimensions), 
thus 
directly relating $\chi_4(t)$ measurements to that of the relevant
lengthscale $\xi_4$ of dynamic heterogeneity.

In experiments, direct 
measurements of $\chi_4(t)$ have been made in
colloidal~\protect\shortcite{weeks2} and granular materials~\protect\shortcite{dauchotbiroli,durian} 
close to the colloidal and granular glass transitions,
and also in foams~\protect\shortcite{mayer} and gels~\protect\shortcite{luca}, 
because dynamics is more easily spatially 
and temporally resolved in those cases.
The results obtained in all these cases are again broadly similar 
to those shown in Fig.~\ref{chi4ludo}, both for the 
time dependence of $\chi_4(t)$ and its evolution upon
a change of the relevant variable controlling the dynamics. 

A major issue is that obtaining information on the behaviour of  
$\chi_4(t)$ and $G_4(r;t)$ from experiments on molecular
systems is difficult. In molecular liquids, it remains a difficult task
to resolve temporally the dynamics at the nanometer scale.
Such measurements are however important 
because numerical simulations and experiments on colloidal
and granular systems can typically only be 
performed for relaxation times spanning at most 5-6 decades.  
On the other hand, in molecular liquids, up to 14 decades can
be measured, and extrapolation of simulation data all the way
to the experimental glass transition is fraught with difficulty.
Indirect estimates of $\chi_4(t)$ from experiments will be discussed
below.

These results are also relevant because many theories of the glass 
transition assume or predict, 
in one way or another, that the dynamics slows down because there are
increasingly large regions over which particles have to relax in a  
correlated or cooperative way, see Sec.~\ref{predictions}
and Ref.~\protect\shortcite{chapter1}.
However, this lengthscale remained elusive
for a long time. Measures of the spatial extent of dynamic heterogeneity, 
in particular $\chi_4(t)$ and   $G_4(r;t)$, seem  
to provide the long-sought evidence of this phenomenon. 
This in turn suggests that the glass transition can indeed 
be considered as a form of critical 
phenomenon involving growing time scales and length scales.
This is an important progress towards the understanding 
of the glass transition phenomenon, even though a clear and 
conclusive understanding of the relationship between dynamic lengthscales 
and relaxation timescales is 
still the focus of intense research. 

\subsubsection{Dependence of $\chi_4(t)$ on the observable
and probe length scale}

As discussed above, one may define a four-point function $G_4(r;t)$ 
starting from any suitable mobility $c({\bf r};t,0)$.  
Indeed, many candidates
have been considered.  It is not our intention to give a detailed
list, but a few comments are in order.

A natural choice for $\chi_4(t)$ is to start from Eq.~(\ref{equ:chi4var}) 
and to take $C(t,0)=\rho_{\bf q}(t)\rho_{-\bf q}(0)$ as the autocorrelation
of a single Fourier component of the density.  In this case,
the average of $C(t,0)$ is the intermediate scattering function
$F({\bf q},t)$ of Eq.~(\ref{isf}).  
In computational studies, it is often more convenient
to construct instead $C(t,0)$ from a simple sum over particles.
That is, one defines the single-particle mobility
\be 
f_i({\bf q},t,0)\equiv e^{i {\bf q} \cdot ({\bf r}_i(t)-{\bf r}_i(0))}, 
\label{equ:fi}
\ee
whose average is the self-part of $F({\bf q},t)$.  The real-space
four-point
function is then given by Eq.~(\ref{equ:g4def}), and the definition
of $\chi_4(t)$ follows. These two definitions of $\chi_4(t)$
are not equivalent. Differences
between them were discussed in 
Ref.~\protect\shortcite{lacevic}, where it was concluded that they contain 
similar information.  

Physically speaking, the key point is that
as particle
$i$ moves away from its initial position $r_i(0)$, the function 
$f_i({\bf q};t,0)$
decays from a value of unity, reaching zero when the particle
has moved a distance of order $(\pi/|{\bf q}|)$.  Once
the particle has moved further than this, the oscillations in the cosine
function imply that averages of $f_i$ give numbers close to zero.
Based on this physical interpretation, other choices for $c_i(t,0)$,
including step functions, or smoothly decaying functions were
used~\protect\shortcite{glotzerfranzparisi,glotzer,lacevic,berthier}. As expected
on physical grounds, constructing four-point functions based on these
choices for $c_i(t,0)$ again leads to 
qualitatively similar behaviours.

Yet another choice is to use a function $c_i(t,0)$ that depends
not just on the positions at time zero and time $t$, but also
on the whole history of the particle between these times.
In particular, one may take a `persistence' function which takes
a value of unity if the particle remains within a distance $a$
of its initial position for all times between 0 and $t$; otherwise
it takes the value zero.  Again, one observes a broadly similar 
behaviour~\protect\shortcite{chandler-chi4}.

All these choices for the definition of the local mobility 
$c_i(t,0)$ involve a `probe length scale',
which is fixed by the choice of measurement, in contrast to the 
sought dynamic length scale $\xi_4(t)$ which should be a 
physical property of the system.
While the specific form of $c({\bf r};t,0)$ is typically unimportant
for the qualitative behaviour of four-point functions, the
quantitative results do depend on the probe length scale.  

Typically, if the probe length scale $a$
is of the order of the particle diameter or smaller,
$c({\bf r};t,0)$ measures local motion, and this is often the scale
on which heterogeneity is most apparent.  As the probe length scale
is increased, contributions to $\chi_4(t)$ come from pairs
of particles that remain correlated over distances comparable
to $a$ and, typically, such correlations weaken as $a$ increases,
reducing $\chi_4(t)$~\protect\shortcite{chandler-chi4}.  Similarly, reducing $a$ also 
reduces $\chi_4(t)$ as short-scale motion corresponding
to thermal vibrations are also typically uncorrelated. 
Therefore, $\chi_4(t)$ is usually maximal for a probe length
scale comparable to the particle size, and it is fixed
to a constant when comparing data at different temperatures 
or densities. An alternative choice is to 
adjust the probe lengthscale $a$ at different state points
such that $\chi_4(t;a)$ reaches its absolute 
maximum, this can be very important for some systems like granular
media close to the rigidity transition where the maximum is reached for values 
of the probe length far below the particle size~\protect\shortcite{lechenaut,lech2,claus}. 
 
\subsubsection{Real-space measurements and
associated structure factors}

We concluded  above that a growing peak in $\chi_4(t)$ 
`directly' reveals the growth of a dynamic correlation lengthscale
as the glass transition is approached. This can only be correct
if the assumptions made above for the scaling form of
$G_4(r;t)$ are correct. Dynamic lengthscales
should in principle be obtained by direct
measurements of a spatial correlation 
function~\protect\shortcite{heuerani,schroder}.   
 
However, in contrast to $\chi_4(t)$, detailed measurements of $G_4(r;t)$
are technically more challenging as dynamic correlations
must now be resolved in space over large distances with a very high
precision, and so there is much less data to draw on.
From the point of view of numerical simulations where 
many measurements of $\chi_4$ were reported, the main limitation 
to properly measure $\xi_4$ is the system size. This might seem surprising 
as typical numbers extracted for the correlation
length scale $\xi_4$ are modest, growing, say, from 1 to at most 5-10
in most reports. However, this modest number hides the fact that
the correlation function only decays to zero for distances
$r$ that are several times larger than $\xi_4$. Given that
$G_4(r;t)$ is accurately measured up to $r = L/2$ in a periodic system of
linear size $L$, going to few $\xi_4$ (say, five times), 
when $\xi_4 \sim 5$ requires 
systems containing at least $N \sim L^3 \sim (2 \times 
5 \times 5)^3 \sim 10^5$ particles in three dimensions, 
assuming the density is near $\rho \approx 1$.  
Such large system sizes are not easily studied 
at low temperatures when relaxation times get very large, even 
with present day computers.
However, these studies are of vital importance in that they allow
the dynamical length scale $\xi_4(t)$ to be measured directly.
Moreover, insights from such studies 
can then be used when inferring the behaviour of
$\xi_4(t)$ from measurements of $\chi_4(t)$.

\begin{figure}[t]
\begin{center}
 \psfig{file=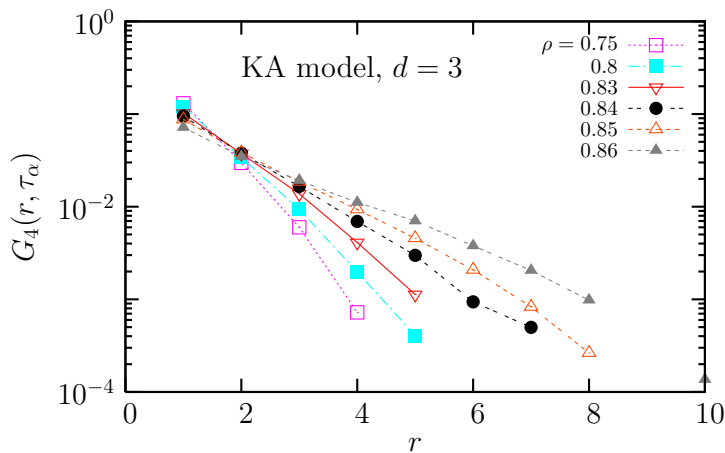,width=9.5cm}
\end{center}
\caption{Four-point correlation function $G_4(r;t=\tau_\alpha)$
measured in computer simulations of the Kob-Andersen
kinetically constrained lattice gas in three dimensions~\protect\shortcite{prlfss}.
The dynamics slows down when density $\rho$ increases, and the 
slower spatial
decay of $G_4$ directly reveals increasingly longer ranged dynamic
correlations accompanying the glass transition.}
\label{fig:g4_example}
\end{figure}

Some representative data are shown in Fig.~\ref{fig:g4_example}.
They are obtained for a lattice gas model with kinetic constraints,
where measurements of $G_4(r;t)$ are somewhat easier than in molecular dynamics
simulations. As discussed above, the idea is that for large $r$,
$G_4(r;t) \simeq A(t) r^{-p} F(r / \xi_4(t))$ which then yields
\be
\chi_4(t) \simeq A(t) \xi_4(t)^{d-p}.
\label{directlink}
\ee
Often, one estimates the prefactor $A(t)$ to be equal
to $G_4(0,t)$, which is simply the variance of the local
quantity $c(r;0,t)$.  However, the accuracy of this estimate
is hard to assess without explicit evaluation of $G_4(r,t)$.
In the example shown in Fig.~\ref{fig:g4_example}, for instance
the scaling between $\chi_4$ and $\xi_4$ in Eq.~(\ref{directlink})
is well obeyed, and 
careful examination of $G_4(r;t)$ suggests that $p \approx 1$
and $A$ is indeed a constant of order 1. 

While this is a subtle situation which requires each case to
be considered individually,
the work in this domain is broadly consistent
with $\chi_4(t)/G_4(0,t)$ representing the number of 
particles involved in heterogeneous relaxation.  Note that these
issues will also be relevant for discussion of other correlations
and susceptibilities in later sections.
Therefore, truly `direct' measurements indeed confirm 
that the increase of the peak of $\chi_4(t)$
corresponds, as expected, to a growing dynamic 
lengthscale $\xi_4(t)$~\protect\shortcite{heuerani,schroder,glotzer,berthier,jcpI}.
As a result of subtleties related to the difference between 
four-point correlations in spin 
glasses and structural glasses, an early study of 
four-point functions~\protect\shortcite{dasgupta-chi4} drew an opposite conclusion, 
but the data of that study are in fact consistent with the 
now-established picture of a growing length scale.

We also note, in passing, that the power $p$ may have more than
one interpretation.  Typically, one assumes that a typical cluster has
size $\xi_4(t)$ and contains $\xi_4(t)^{d-p}$ particles,
so that $d-p$ is interpreted as a fractal dimension.  However,
an alternative would be that clusters are all compact, but that
the distribution of their sizes is rather broad.  This uncertainty
reflects the fact that four-point functions involve 
averages over many clusters, so that they do not resolve details
of cluster structure.

Instead of direct inspection of $G_4(r;t)$, it is often
convenient to consider its Fourier transform
$S_4(k;t) = \int\mathrm{d}^dr e^{i {\bf k} \cdot {\bf r}} 
G_4(r;t)$~\protect\shortcite{glotzer,poole2,glotzer3,schroder,berthier,yamamoto1,yamamoto2,flenner-szamel}.
In particular, this allows data for different times
and different temperatures to be combined into a scaling analysis
that can yield the temperature dependence of $\xi_4$, leaving
an uncertain prefactor.  This approach was taken for example
in Refs.~\protect\shortcite{yamamoto2,jcpI,flenner-szamel}. 
Typically, simulation data result
in length scales between 1 and 5 diameters.
 
So far, we have considered circularly-averaged $G_4(r;t)$
and $S_4(k;t)$.
However, if the probe function $c_i(t,0)$ is anisotropic, as in
Eq.~(\ref{equ:fi}), four-point functions $G_4(r;t)$ 
then also depend on the angle
between the separation vector ${\bf r}$ and the probe vector ${\bf q}$.
Several papers~\protect\shortcite{weeks2,szamel-aniso,heuerani} 
have investigated this issue and found
that motion in longitudinal directions is indeed more strongly 
correlated than motion in transverse directions, such that
$G_4(r;t)$ is truly an anisotropic function. These findings 
add further to the difficulty of extracting the length scale
$\xi_4$ from direct measurements of four-point structure
factors.

\subsection{Experimental estimates of multi-point susceptibilities} 
\label{inequality}

Although readily accessible in numerical simulations, the fluctuations of $C(t,0)$ that give 
access to $\chi_4(t)$ are in general very small and impossible to measure directly in 
experiments, except when the range of the dynamic correlation is 
macroscopic, as in granular materials~\protect\shortcite{dauchot} or in
soft glassy materials where 
it can reach the micrometer and 
even millimetre range~\protect\shortcite{mayer,luca}.
To access $\chi_4(t)$ in molecular liquids, one should
perform time-resolved dynamic measurements probing very small volumes,
with a linear size of the order of a few nanometers.  Although possible, 
such experiments remain to be performed with the required 
accuracy.

It was recently realized that 
simpler alternative procedures exist. The central idea underpinning
these results is that induced dynamic fluctuations are 
in general more easily accessible than spontaneous 
ones, and both types of fluctuations 
can be related to one another by fluctuation-dissipation 
theorems. The physical motivation is that while four-point 
correlations offer a direct probe of the dynamic heterogeneities, 
other multi-point correlation functions give very useful information 
about the microscopic mechanisms leading to these heterogeneities. 

For example, one might expect that the slow part of a local enthalpy 
(or energy, density) fluctuation $\delta h_x(t=0)$ at position $x$ 
and time $t=0$ triggers or eases the dynamics in its surroundings, 
leading to a systematic correlation between 
$\delta h_x(t=0)$ and $c(x+r;t,0)$. This 
physical intuition suggests the definition of a family of 
three-point correlation functions that relate thermodynamic or 
structural fluctuations to dynamical ones. Interestingly, 
and crucially, some of these 
three-point correlations are both experimentally accessible and give 
bounds or approximations to the four-point dynamic 
correlations\protect\shortcite{science,jcpI,jcpII}. 

Based on this insight, one may obtain a lower bound 
on $\chi_4(t)$ using the Cauchy-Schwartz inequality 
$\left\langle \delta H(0)\delta C(t,0) \right\rangle^{2} \leq \left\langle
  \delta H(0)^{2}\right\rangle \left\langle \delta C(t,0)^{2}\right\rangle$,
where $H(t)$ denotes the enthalpy at time $t$, and 
$\delta X = X - \langle X \rangle$ denotes the fluctuating 
part of the observable $X$. 
By using a fluctuation-dissipation relation valid when the energy
is conserved by the dynamics, the previous inequality 
can be rewritten as~\protect\shortcite{science}: 
\begin{equation}
\label{eq7:equation}
\chi_{4}(t) \geq \frac{k_{B}T^{2}}{c_{P}} \left[ \chi_{T}(t) 
\right]^{2},
\end{equation}
where the multi-point response function $\chi_T(t)$ is defined by 
\begin{equation}
\chi_T(t)= \frac{\partial \langle C(t,0) \rangle}{\partial T}
\bigg\vert_{N,P}=\frac{N}{k_{B}T^{2}}\left\langle \delta H(0)\delta
  C(t,0)\right\rangle,
\label{cov} 
\end{equation}
and $c_P$ is the specific heat per particle (at constant pressure).

In this way, the experimentally accessible response
$\chi_{T}(t)$ which quantifies the sensitivity of average
correlation functions $\langle C(t,0) \rangle$ 
to an infinitesimal temperature change, 
can be used in Eq.~(\ref{eq7:equation}) to yield 
a lower bound on $\chi_{4}(t)$.
From Eq.~(\ref{cov}), it is clear that $\chi_T$ is directly
related to the covariance of enthalpy and dynamic fluctuations, 
and thus captures the part of dynamic heterogeneity which is 
triggered by enthalpy fluctuations.  
 
Detailed numerical simulations and 
theoretical arguments~\protect\shortcite{jcpI,jcpII} 
strongly suggest that the right hand side of (\ref{eq7:equation}) 
actually provides a good estimate of $\chi_4(t)$ in supercooled liquids, 
and not just a lower bound.
Similar estimates exist considering density as a perturbing field 
instead of the temperature. These are useful when considering 
colloidal or granular
materials where the glass transition is mostly controlled by the 
packing fraction. The quality of the corresponding lower bound was 
tested experimentally 
on granular packings close to the jamming 
transition \protect\shortcite{lech2}, and numerically
for colloidal hard spheres~\protect\shortcite{gio}. 

\begin{figure}[t]
\begin{center}
 \psfig{file=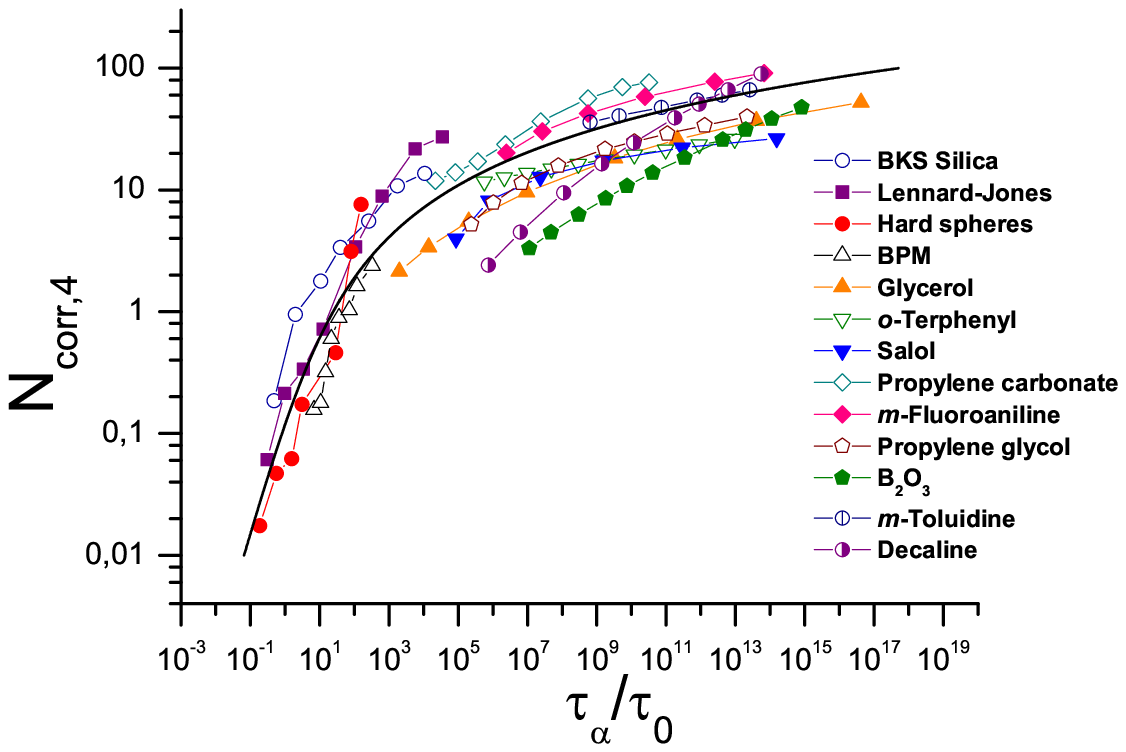,width=10.5cm}
\end{center}
\caption{\label{cecile} 
Dynamic scaling relation 
between the number of dynamically correlated particles, $N_{{\rm corr},4}$, 
and relaxation timescale, $\tau_\alpha$, for a number of 
glass-formers~\protect\shortcite{cecile}, determined using the 
bound provided by Eq.~(\ref{eq7:equation}). For all systems, dynamic 
fluctuations increase when the glass transition is 
approached. The full line through the data~\protect\shortcite{cecile} 
suggests a crossover from 
algebraic, $N_{{\rm corr},4} \sim \tau_\alpha^z$,
to logarithmic, $N_{{\rm corr},4} \sim \exp (\tau_\alpha^\psi)$,
growth of dynamic correlations with increasing $\tau_\alpha$.} 
\end{figure}

Using this method, Dalle-Ferrier {\it et al.}~\protect\shortcite{cecile} have been able to 
estimate the evolution of the peak value of $\chi_4$ 
for many different glass-formers in the entire supercooled
regime. In Fig. \ref{cecile} we show some of these results as a 
function of the relaxation
timescale. The value on the
$y$-axis, a bound on the peak of $\chi_4$, is a proxy for the number of 
molecules, $N_{{\rm corr},4}$ 
in a cluster of mobile or immobile particles.  As discussed briefly
above,
$\chi_4(t)$ is expected to be equal to $N_{{\rm corr},4}$, up to a
proportionality constant $A(t)$ which is not known
from experiments, probably explaining why
the high temperature values of $N_{{\rm corr},4}$ are smaller than one. 
Figure~\ref{cecile} also indicates that $N_{{\rm corr},4}$ grows faster 
when $\tau_{\alpha}$ is not very large, 
close to the onset of slow dynamics, and 
a power law relationship between $N_{{\rm corr},4}$ and $\tau_\alpha$ 
is good in this regime ($\tau_\alpha /\tau_0 < 10^4$). 
The growth of $N_{{\rm corr},4}$ with $\tau_\alpha$ 
becomes much slower closer to the glass transition 
temperature $T_{g}$, where
a change of 6 decades in time corresponds to a mere increase of a
factor about 4 of $N_{{\rm corr},4}$, suggesting logarithmic
rather than power law growth of dynamic correlations.
A similar crossover towards a very slow growth 
of dynamic correlations is reported for 
colloidal hard spheres~\protect\shortcite{gio} and model glasses~\protect\shortcite{jcpI}, 
and is observed in numerical 
simulations even if the dynamic lengthscale $\xi_4$
is directly estimated~\protect\shortcite{flenner-szamel}.   
The consequences of such an effect for theories of the glass
transition are discussed below.
Bearing in mind all the caveats discussed above (unknown prefactors, 
quality of the bound, etc.), the experimental data compiled in
Fig.~\ref{cecile} do appear to confirm that 
dynamic fluctuations and correlation lengthscales do grow
when the molecular liquids approach their glass transitions.

\subsection{Four-point susceptibilities: some caveats} 
\label{caveats}

The above story about the four-point susceptibility looks quite enticing. 
We have essentially argued
that dynamical heterogeneity should be quantified by the spatial 
correlations of the mobility. This
correlation function is a priori hard to measure in molecular glasses, 
but a divine surprise occurs:
using rather trivial mathematics, its spatial integral is found 
to be bounded from below by a quantity that 
is much easier to measure. 
Is this too good to be true? What is the physics 
underpinning this `easy' bound?

The answer is that four-point correlation functions pick up a contribution that depends both on the statistical ensemble used (i.e. NVE vs. NVT)
and on the dynamics (i.e. Brownian vs. Newtonian). 
Using general scaling arguments based on a dominant length scale
$\xi_4$, one can show that in Fourier space the 
four-point correlation function has the following 
structure (in the $k \to 0$ limit)~\protect\shortcite{jcpI,jcpII}:
\begin{equation}
S_4(k;\tau_\alpha) = \xi_4^s \hat H_1(k\xi_4) + g(k) \left[\xi_4^s \hat H_2(k\xi_4)\right]^2,
\end{equation}
where $s$ is a certain exponent (related to $p$ above through $2s=d-p$) and $\hat H_{1,2}$ are certain scaling functions which behave
similarly at small and large arguments. The wavevector dependent function $g(k)$ carries the dependence on the statistical ensemble and dynamics. 
In particular, $g(k=0)=0$ when all conserved quantities are strictly fixed, as in the NVE ensemble\footnote{Actually, fast degrees of freedom can give subleading contributions in some cases.}. Similarly, $g(k)$ is very different for Brownian 
dynamics (for which energy is not conserved) or for Newtonian dynamics (that conserves energy). So one has to face the uneasy truth that $\chi_4(t) \equiv S_4(k=0;t)$ 
depends on many microscopic and macroscopic details, although physically these should not affect the dynamic correlation length $\xi_4$. It is definitely not so easy to directly relate
$\chi_4$ to a dynamic correlation length. Intuitively, conserved variables play a role in ``transmitting'' the information about mobility from
one region to another. 

From the above expression, one sees that $S_4(k;t)$ mixes up two contributions that one would like to disentangle so as to extract the relevant scaling contribution 
from the first term only. It turns out that this first term is proportional to a three-point response function, that measures the change of the dynamics induced 
by a perturbation some distance away, and that we will discuss in more 
detail in section~\ref{beyond}. 
It is this three-point response, not $G_4$, that is the fundamental 
object carrying information about dynamical heterogeneities, and from which $G_4$ is constructed. The space integral of the three-point response function is $\xi_4^s \hat H_1(0)$;
physically it represents the response of the dynamics to a uniform shift of an external parameter, such as the temperature or the density. Hence $\chi_T(t)$ as defined 
in the previous section is a three-point function at 
zero wavevector. This explains the physical nature of the lower bound on $\chi_4(t)$: due to the contribution of the energy as a conserved quantity, $\chi_4$ has  
two contributions: one proportional to $\chi_T$ and one proportional to $\chi_T^2$, the latter being precisely the lower bound of the previous section.

This discussion leads to the following caveats, that we alluded to above: a)~the identification of a correlation volume from $\chi_4(t)$ alone is not warranted
in general. The information contained in $S_4(k;\tau_\alpha)$ is needed to 
unambiguously relate the growth of $\chi_4(t)$ to a growing 
lengthscale;\footnote{Related to this point, it is worth mentioning the case of 
purely Arrhenius systems, which are considered to be non-cooperative systems. Still, the lower bound based on $\chi_T$ proves that $\chi_4(\tau_\alpha)$ diverges in these
systems as least as $T^{-2}$ as the temperature goes to zero~\protect\shortcite{cecile}. 
The physical interpretation of such an apparent growth of the range of dynamical correlations is still unclear.} 
b)~extracting information from $S_4$ can be difficult due to 
interference effects between the two terms and c)~three-point response functions are the fundamental building bricks 
for dynamic correlations, and are not soiled with problems related to conserved variables or statistical ensembles. 

\section{Theoretical discussion}

\subsection{Recent progress based on four-point functions}

In the previous section, we have summarised some of the properties
of four-point functions, their advantages for calculating the extent
of dynamical heterogeneity, and some direct and indirect measurements
of these quantities.

There are many subtleties associated with these measurements, but the
same broad picture is observed in a variety of systems and
is robust to the precise measurement used.  Essentially, as relaxation
times increase, the four-point susceptibility increases, suggesting 
the presence of a growing dynamic correlation length.  
Where the  real-space
function $G_4(r;t)$ can also be measured, this confirms 
more directly the increase
of such a length scale, typically in the range between 1 and 10 molecular
diameters.  The fundamental, unavoidable conclusion seems to 
be that glassy behaviour is not a purely local `caging' 
of particles by their neighbours, but indeed a genuine 
collective phenomenon.  

Having established the existence of a growing correlation length,
several questions arise. From a theoretical perspective, we are
familiar with the idea, borrowed from equilibrium
critical phenomena, that when correlation length scales get large,
microscopic features of the system become unimportant, and 
`universal' behaviour emerges. Whether realistic glassy systems
have length scales that are large enough for such a universal
description remains unclear.  Many analyses in this spirit
have nevertheless been attempted, as we shall discuss shortly.
It is likely that to in order to reach a good quantitative agreement a careful
treatment of pre-asymptotic effects will have to be performed.

A second fundamental point concerns the microscopic mechanisms that give
rise to the correlations revealed by four-point functions.
Many model systems can demonstrate the presence
of increasing time scales, coupled with increasing susceptibilities
$\chi_4(t)$ and length scales $\xi_4$.  Predictions from different
theoretical frameworks of the form of four-point functions
are discussed in the next section, and we evaluate some of the 
theories in the light of existing results.

\subsection{Models of the glass transition and their 
predictions of dynamic heterogeneity}

\label{predictions}

We now turn to  perhaps the most fundamental
question in this area: what are the dominant mechanisms by which
structural relaxation takes place in glassy materials?  
We give a quick survey of the dominant pictures for molecular
glasses, their predictions for four-point functions, and the 
extent to which these are borne out.

\subsubsection{Mode-coupling theory}

The mode-coupling theory (MCT) of the glass transition~\protect\shortcite{mct}
was historically derived from liquid state theory. Starting from
exact microscopic equations of motion for the density field 
in a liquid, several uncontrolled approximations are then
performed to yield a closed set of dynamical equations 
for intermediate scattering functions. These equations 
give rise to a dynamic singularity at some finite 
temperature, $T_c$, where relaxation times
diverge in an algebraic manner. Additionally, 
very precise quantitative predictions
can be made about the specific form of intermediate scattering 
functions, suggesting a very rich behaviour of time 
correlation functions, which do resemble the 
behaviour observed experimentally and reported in Fig.~\ref{fqt}.
As is well-known these 
predictions only apply over a modest time window of about 2-3 decades
in the moderately supercooled regime, but dramatically
break down nearer to the glass transition~\protect\shortcite{mctbook}. 
 
Another dramatic failure of the traditional formulation of the theory, 
more relevant to the present contribution, is its  
inability to accurately predict the shape of the van-Hove 
distribution function described  above, the resulting wavevector
dependence of the self-intermediate scattering 
function (especially at low wavector), and the 
corresponding decoupling between self-diffusion 
constant and the viscosity. 

Following this historical route, therefore, 
it is not obvious whether the MCT dynamic singularity is accompanied 
by non-trivial dynamic fluctuations. This also means that the theory is 
not easily interpreted in physical terms. 
Recently, mode-coupling theory was reformulated in such a way 
that both these issues were greatly clarified. Using a
field-theoretic formulation, it is possible to 
perform consistent mode-coupling approximations to get analytical 
predictions for both averaged two-time dynamic correlation
functions and for the dynamic fluctuations around the averaged
behaviour, i.e. for $\chi_4(t)$~\protect\shortcite{franzparisi,jcpII,bbepl}. 
In a subsequent recent move,
`inhomogeneous mode-coupling' predictions for the shape 
and scaling form of four-point spatial correlation functions $G_4(r;t)$ and 
its Fourier transform $S_4(k;t)$ were finally obtained~\protect\shortcite{IMCT}. Thus, 
overall, mode-coupling theory is now able to make an impressive set 
of very detailed predictions for a very large family of spatio-temporal
correlation functions for any given liquid, starting from 
the form of the microscopic interaction between the particles.   

A few numerical simulations have been presented to test these
new predictions. First, the temporal evolution 
of the four-point susceptibility $\chi_4(t)$ was compared to
mode-coupling predictions. Just as time-correlation
functions decay within MCT in a two-step process
similar to the data presented in Fig.~\ref{fqt}, $\chi_4(t)$ 
is predicted to grow with time with two distinct power laws, 
$\chi_4 \sim t^a$ and $\chi_4 \sim t^b$ in the time regimes
respectively corresponding to the approach to, and departure from, 
the plateau. These two power law regimes have been 
successfully identified in numerical work, with numerical 
values for the exponents $a$ and $b$ that are in `reasonable'
agreement with numbers predicted by MCT~\protect\shortcite{jcpII,berthiersilica,lj-mc}. 

The peak of the 
four-point susceptibility is predicted to diverge algebraically
at the critical temperature. This prediction
was observed numerically to hold over a similar (restricted) 
temperature window as for the averaged relaxation time $\tau_\alpha$
itself. This finding implies that the peak of $\chi_4$, when plotted
as a function of $\tau_\alpha$, follows a power law
scaling, $\chi_4 \sim \tau_\alpha^z$, where $z$ is 
predicted to be a non-universal critical exponent.
Returning to the data compilation in Fig.~\ref{cecile}, we 
remark that the data obtained in the moderately supercooled
regime do indeed approximately follow an initial growth
which is  consistent with the MCT prediction, while 
clearly breaking down at lower temperatures.  

Finally, predictions for the detailed shape 
of four-point correlation functions, in particular 
for $S_4(k;t)$, were recently confronted
to numerical results, with inconclusive results.
While a first paper~\protect\shortcite{andersenstein} 
reports excellent agreement with MCT predictions 
both for the wavevector dependence of $S_4(k;t)$ and its evolution 
with temperature, a more recent report~\protect\shortcite{karmakar} 
claims that disagreements with theoretical predictions 
arise when larger system sizes are included in the 
numerical analysis. This ongoing debate illustrates the fact
mentioned above that even for the modest correlation
length scales characterising relaxation in supercooled liquids, 
very large system sizes are needed to unambiguously 
and accurately measure four-point functions. Clearly, more 
work is needed to clarify the status of the
large body of MCT predictions regarding four-point functions.

\subsubsection{Facilitation picture and kinetically constrained models}

In the facilitation picture of supercooled liquids, structural relaxation is 
thought to originate from propagation of 
localised excitations, or `defects', throughout the system~\protect\shortcite{glarum}. 
The idea is that the local structure at position $x$ 
changes rapidly when a defect visits the neighbourhood of $x$. 
Then, one makes the further hypothesis
that defects are sparse, and most
of the system is in fact immobile, such that defects can be 
considered as uncorrelated, independent objects.
In practice, very little is known about the nature, origin,
or even the existence of such defects in real liquids.  

Nevertheless, it is clear within this picture that dynamics is highly 
heterogeneous
in space, and temporally intermittent. Physically, one
expects a dynamic correlation lengthscale $\xi_4$
to emerge, which should basically correspond to the linear
size of the region explored independently
by a given defect. This also means 
that the time dependence of $\xi_4(t)$, and thus of $\chi_4(t)$
can directly be connected, in this view, to how fast the defects
move. In the simplest approximation where defects are diffusive objects,
one would expect $\xi_4(t) \sim \sqrt{t}$ for $t < \tau_\alpha$. 

In recent years, these ideas have been pursued quite extensively,
based on the proposal~\protect\shortcite{GC-pnas} that facilitation is
essential for explaining the heterogeneous dynamics of supercooled 
liquids.  The theory has been developed primarily
through studies of systems called kinetically constrained 
models\protect\shortcite{Ritort-Sollich}.  Numerous distinct lattice models
belong to this family, which can be distinguished by the set
of microscopic rules governing the dynamics of the localised 
defects, the existence or absence of conservation laws, 
the topology of the lattice, etc.
The simplest models, such as the (one-spin facilitated) 
Fredrickson-Andersen (FA) model, in fact reproduce nearly exactly the 
scaling relations mentioned above for four-point 
functions~\protect\shortcite{TWBBB,chandler-chi4,steve,steve2}. 

Given the large number of distinct models, and the fact that 
models are postulated instead of being derived as approximate,
coarse-grained representations of liquids, it is not clear how
one should compare their behaviour to numerical results 
obtained for realistic liquid models. 
This issue is discussed in a recent review~\protect\shortcite{annu}.
Thus, it is perhaps 
better to interpret this diversity as being suggestive of the different
types of behaviour one can possibly encounter in liquids.
On the other hand, from the theoretical point of view, having 
well-defined, relatively simple, statistical models defined on the
lattice is very appealing as very many detailed and quantitative  
results can be obtained by exploiting tools from statistical mechanics.  
Thus, kinetically constrained models can also be viewed as 
`toy supercooled liquids'.
  
In this regard, the study of the dynamically 
heterogeneous behaviour of kinetically constrained models 
has been a very active field of research in recent years. 
Many models have been investigated and a large number
of time correlation functions (two-point, four-point, 
persistence functions) have been analysed, suggesting possible 
behaviours for dynamic susceptibilities~\protect\shortcite{jcpI,chandler-chi4} or 
decoupling 
phenomena~\protect\shortcite{jung,pan}. 
We refer to the chapter by Sollich, Toninelli, and Garrahan for 
further details and references on this topic. 

At the qualitative level, it is obvious that all models are 
characterised by rapidly growing time scales and length scales, 
and are thus interesting models to study dynamic heterogeneity. 
However, models with 
diffusive point defects (like the simplest 
of Fredrickson-Andersen models), 
do not compare well with the real liquids that have been studied so far. 
In three dimensions, 
they predict simple exponential relaxation and no decoupling  
phenomena~\protect\shortcite{jack-mayer-sollich}. The dependence of $\chi_4$
on time and temperature are also characterised by scaling laws that
have not been observed in numerical and experimental results~\protect\shortcite{TWBBB,science}.
However, the Arrhenius scaling of their relaxation time indicates a relation
between these models and
`strong' liquids~\protect\shortcite{GC-pnas,steve2} and there are comparatively few results
on dynamical heterogeneity for such materials.  In particular, while such
models predict rather large dynamical length scales in strong materials~\protect\shortcite{GC-pnas,nef}, 
these have not yet been observed.

On the other hand, more complicated models
where defects move sub-diffusively or cooperatively
seem to be more appropriate representations of `fragile' liquids, which have a 
non-Arrhenius scaling of relaxation time with temperature.
Such kinetically constrained models exhibit stretched exponential relaxation
as in Fig.~\ref{fqt}, decoupling phenomena~\protect\shortcite{jung} similar to the 
results in Fig.~\ref{otp}, realistic form of four-point structure
factors as in Fig.~\ref{fig:g4_example}, or
dynamic length scales which grow very slowly~\protect\shortcite{GC-pnas},
in qualitative agreement with 
the experimental results shown in Fig.~\ref{cecile}.

\subsubsection{Adam-Gibbs and the mosaic picture}
\label{rfot}

The idea that relaxation events in glasses are collective and involve the simultaneous motion of several particles dates 
back at least to Adam and Gibbs, who provided an argument to relate the size of these ``cooperatively rearranging regions'' (CRR)
to the configurational entropy of the supercooled liquid. A lot of the work on dynamical heterogeneities is in fact motivated
by the Adam-Gibbs picture and attempts to determine the size of 
these CRR~\protect\shortcite{binderkob}. 

The Adam-Gibbs picture was later put on more  solid ground in the context of the Random First Order Transition (RFOT) Theory of glasses~\protect\shortcite{KTW,rfot}. 
RFOT suggests that supercooled liquids be can be thought of as a mosaic of ``glass nodules'' or ``glassites'' with a spatial extension
$\ell^*(T)$ limited by the configurational entropy. Regions of size smaller than $\ell^*(T)$ are ideal glasses: they cannot relax, even on very long time scales,
because the number of states towards which the system can escape is too small to compete with the energy that blocks the system in a given favourable 
configuration. Regions of size greater than $\ell^*$ are liquid in the sense that they explore with time an exponentially large number of unrelated configurations, and
all correlation functions go to zero. The relaxation time of the whole liquid is therefore the relaxation time of glassites of size $\ell^*$. This relaxation occurs 
through collective activated events that sweep a region of size $\ell^*$, 
which are the CRR regions of the Adam-Gibbs theory. 
The crucial assumption of RFOT is that 
thermodynamics alone fixes the value of  $\ell^*$, whereas the relaxation time $\tau_\alpha$ involves the 
height of the activation barriers on scale $\ell^*$, which is assumed to grow as a power-law, $\ell^{*\psi}$, where $\psi$ is a certain exponent~\protect\shortcite{bbpsi}. 

Note that when an activated event takes place within a glassite of size $\ell^*$, the boundary conditions of the
nearby region changes. There is a substantial probability that this triggers, or facilitates, an activated
event there as well, possibly inducing an ``avalanche'' process that extends over the dynamic correlation length scale $\xi > \ell^*$. 
The dynamics on length scales less than $ \ell^*$ is, within RFOT, inherently cooperative, but the relation between the dynamic correlation length $\xi$, defined for example through 
three- or four-point point correlation function and the mosaic length $\ell^*$ is at this stage an important open problem (see \protect\shortcite{cecile,zamponi}). 
If $\xi$ is of the order of a few glassite lengths $\ell^*$, then one expects that $\chi_4(\tau_\alpha)$ should grow as $\ell^*$ to some power. Assuming activated 
scaling, $\ln \tau_\alpha \sim \ell^{*\psi}$, finally leads to $\chi_4 \sim (\ln \tau_\alpha)^z$, instead of a power-law relation predicted by MCT or by non-cooperative KCMs.
The crossover towards this logarithmic behaviour is not incompatible with the data~\protect\shortcite{xia}, see section \ref{inequality} above. In fact, the details of the crossover between the MCT region and 
the RFOT region are still very mysterious 
(see \protect\shortcite{reviewrfot} for a recent discussion), but some claims have been made about the evolution of the shape of the dynamically 
correlated regions, that should morph from stringy, fractal objects in 
the MCT region to compact blobs at lower temperatures~\protect\shortcite{stringrfot}. 
It would be interesting to 
devise some experimental protocol to test these predictions. 

\section{Beyond four-point functions: other tools to 
detect dynamical correlations}
\label{beyond}

So far, we have discussed how four-point functions can
be used to estimate dynamical length scales, and we have stated that
these are typically found to be in the range 1-10 molecular
diameters.  However, we have also noted that (a) the four-point susceptibility estimate of the 
dynamical correlation volume may lead to erroneous results (see section \ref{caveats}) and (b) a variety of
different theories of the glass transition are broadly consistent with the above estimate.  

To make further progress, it seems that more adapted and discriminating observables will be required.
In fact, there are a wealth of methods that have been used to characterise
dynamical heterogeneity, of which we discuss just a few, and we refer
to other chapters in this book for details.  
Here, we mainly emphasise the questions
that can be addressed by different methods, and give
an overview of the relationships
between some of the methods that have been 
developed in different contexts.

\subsection{Non-linear susceptibilities}

In standard critical phenomena, diverging two-point correlations lead to singular linear responses. 
It is therefore quite natural to conjecture that increasing dynamical correlations should also lead to anomalous 
responses of some kind. Spin glass theory provides, again, an interesting insight. As discussed in section \ref{spin glass perspective} 
the spin glass transition (at zero external field) is signaled by the divergence of the four-point static correlation function 
$\chi_{SG}$. It can be easily established that close to the transtion, $\chi_{SG}$ is related to the third-order non-linear magnetic 
response at zero frequency $\chi_{3}(\omega=0)$ \protect\shortcite{Hertz}:
\be
\chi_{3}(\omega=0) = -\frac{\chi_{SG}-2/3}{ (k_B T)^2 }.
\ee
Thus, although linear responses are blind to the development of spin glass long range order, the non-linear magnetic response is not.
Actually, it diverges at the transition and, hence, is a direct experimental probe, contrary to   $\chi_{SG}$, which can 
instead only be measured in numerical simulations.

The analogy with spin glasses discussed in section \ref{spin glass perspective} therefore suggests
that glasses should also display increasingly non-linear responses approaching the glass transition, as first argued in 
\protect\shortcite{BBPRB}. 
Theoretically, this can be 
substantiated by some general scaling arguments and by 
a mode-coupling calculation.
These are described in \protect\shortcite{nonlinearJCP}; we will just briefly summarize them in the following. The starting point
is to rewrite the generic third order non-linear response $\chi_3(t)$ in terms of the second order change $R_2$ of the linear response $R$:
\begin{equation} \label{eqr2}
\chi_3(t)=\int^t_{-\infty} dt_1 dt_2 dt_3 \frac{\delta P(t)}{\delta E(t_1)\delta E(t_3) \delta E(t_2)}E(t_1)E(t_2)E(t_3)=
\end{equation}
\[
\int ^t_{-\infty} dt_1 dt_2 dt_3E(t_1)\frac{\delta R(t,t_1)}{\delta E(t_2) \delta E(t_2)}E(t_2)E(t_3)=\int^t_{-\infty}dt_1 E(t_1)R_2(t,t_1)
\]
Note that we will focus on the dielectric non-linear response (so $P$ is the electric polarization) but the generalization to other perturbing field is straightforward\footnote{Depending on the perturbing field, the symmetry $E\rightarrow -E$ will hold or not. In the latter case, the first non-linear response is the quadratic one, in the former the quadratic vanishes by symmetry and one has to focus on the third order one.}. 
It is easy to understand, at least at low frequency, why $R_2$ and therefore the non-linear susceptibility 
have a singular behavior. In fact, within an adiabatic approximation, one finds that the linear response 
in the steady state created by a slowly alternating field is:
\be \label{eq:sl}
R_{eq} (t - t', E \cos (\omega t)),
\ee
where $R_{eq} (t - t', E)$ is the equilibrium response function with a {\it static} field $E$.
Since we are interested in the small $E$ behavior, 
we can expand the above expression up to second
order in $E$, this yields:
\begin{equation} \label{eq:lfl}
R_2(t,t_1)=R_{eq}(\tau, E \cos (\omega t))- R_0 (\tau )\approx \frac{E^2 \cos^2(\omega t)}{2} \, \frac{\partial^2 R_{eq}(t-t_1,E)}{\partial E^2} \bigg |_{E=0},
\end{equation}
where $R_0 (\tau)$ is the unperturbed equilibrium response function, 
and the derivative is computed with respect to a constant external field.
The second term is expected to give a singular contribution because close to the glass transition
a small applied field $E$ is roughly equivalent to a 
shift of the order of $E^2$ of the glass transition 
temperature (see below), and $\tau_\alpha$ significantly 
varies when the temperature is changed by a small amount close to $T_g$.
More precisely, by taking into account the $E\rightarrow -E$ symmetry one can rewrite $\partial^2 R_{eq}/\partial E^2$ as $2 \partial R_{eq}/\partial \Theta$ where 
$\theta=E^2$. Using the time temperature superposition, one finds:
\begin{equation} 
\frac{\partial R_{eq}(\tau,E)}{\partial \Theta} \bigg |_{E=0}\simeq \kappa\frac{\partial R_{eq}(\tau,0)}{\partial T}.
\end{equation}
where $\kappa=\frac{\partial \tau_\alpha/\partial \Theta}{\partial \tau_\alpha/\partial T} $. 
Using that $R_{eq}(\tau,0)$ is the Fourier transform of the linear susceptibility $\chi_1(\omega)$ and plugging 
the previous expressions in (\ref{eqr2}) (and after some algebra detailed in \protect\shortcite{nonlinearJCP}) one finds 
the following result: 
\begin{equation} \label{eq:chi3bis}
\chi_3(\omega) \approx  \kappa \frac{\partial \chi_1(2 \omega)}{\partial T},
\label{link}
\end{equation}
which is expected to hold at low enough frequency, 
at least when the deviations from time temperature 
superposition are weak. $\kappa$ is expected to be a slowly varying 
function of temperature, a constant in first 
approximation~\protect\shortcite{nonlinearJCP}.
In this expression, $\partial \chi_1(\omega) / \partial T$
is akin to the three-point susceptibility 
$\chi_T$ defined in section \ref{inequality}.
Thus, Eq.~(\ref{link}) is an important result since it establishes a 
relationship with the linear dynamical responses that 
have been used to evaluate dynamical correlations, and 
it also proves that supercooled liquids should respond
in an increasingly non-linear way approaching the glass transition since, 
as we have discussed before, $\chi_T$ and therefore $\frac{\partial \chi_1(2 \omega)}{\partial T}$
increase approaching the glass transition. 

\begin{figure}[t]
\begin{center}
 \psfig{file=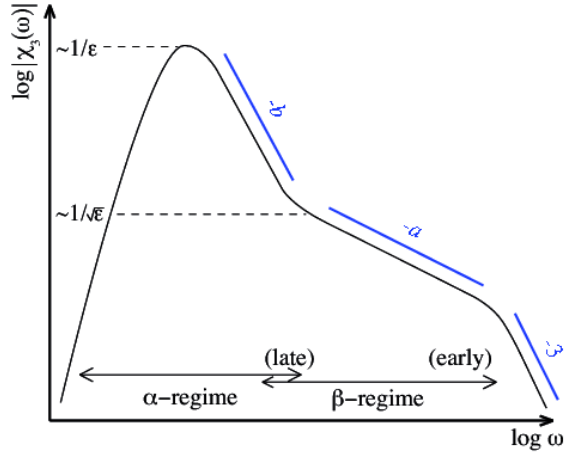,width=7.5cm}
\end{center}
\caption{Sketch of $\log |\chi_3 (\omega)|$ as a function of
$\log \omega$, showing different frequency 
regimes and crossovers~\protect\shortcite{nonlinearJCP}: $\omega
\tau_\alpha\ll 1$, $\omega \tau_\alpha \sim 1$,
$\tau_\beta/\tau_\alpha\ll\omega \tau_\beta \ll 1$ ($\epsilon=T-T_c$), 
$\omega \tau_\beta \gg 1$, $\omega \tau_0 \sim 1$.
Note that the low frequency limit is 
non zero but much smaller than the peak value for $T$ close to $T_c$.}
\label{fig:chi3}
\vspace{0.0cm}
\end{figure}

The above general heuristic arguments can be supplemented by more microscopic ones based on MCT, which provides quantitative predictions on the critical behaviour of 
$\chi_3$. Although the corresponding results are restricted to the small temperature regime where MCT is believed to apply,
they are nevertheless guidelines for the general behaviour of  $\chi_3$. 
We sketch the evolution of the absolute value of  $\chi_3$ with frequency in Fig.~\ref{fig:chi3}. 

\begin{itemize}
\item In the $\alpha$-regime, i.e. $\omega\sim 1/\tau_\alpha\sim\epsilon^{1/2a+1/2b}/\tau_0$, the absolute value of $\chi_3(\omega)$ grows with decreasing $\omega$ and reaches its maximum, of height of order $1/\epsilon$, after which it decreases as $\omega^{-b}\tau_\beta$ at large $\omega$. In this regime, one has the scaling form: $\chi_3(\omega)=\frac{1}{\epsilon}{\mathcal G}(\omega\tau_\alpha)$.

\item At the crossover between the early $\alpha$-regime and late $\beta$-regime the absolute value of $\chi_3(\omega)$ is of order $1/\sqrt{\epsilon}$.

\item In the $\beta$-regime, i.e. $\omega\sim 1/\tau_\beta\sim\epsilon^{1/2a}/\tau_0$, the absolute value of $\chi_3(\omega)$ decreases as $\omega^{-b}\tau_\beta$ at small $\omega$ and as $\omega^{-a}\tau_\beta$ at large $\omega$. In this regime
one has a scaling form $\chi_3(\omega)=\frac{1}{\sqrt{\epsilon}}{\mathcal F}(\omega\tau_\beta)$
\end{itemize}

Exponents $a$ and $b$ are the well known critical exponents of MCT
introduced above; $\tau_0$ is a microscopic relaxation time and $\epsilon 
= (T - T_c)/T_c$ is the distance from the mode-coupling critical temperature $T_c$. We remark that the existence of the peak and the decrease at low frequency is a non trivial prediction since  it is in contrast with the (trivial) non-linear response of uncorrelated Brownian dipoles~\protect\shortcite{perpignan} and with spin glasses. The decrease with an exponent three at high frequency 
sketched in Fig.~\ref{fig:chi3} is instead trivial and present also for independent dipoles \protect\shortcite{perpignan}. 

Recent experiments of third-order non-linear dielectric 
responses of supercooled glycerol have indeed shown, for the first time, that
these theoretical expectations are qualitatively correct \protect\shortcite{chi3preprint}. We refer to the chapter by Alba-Simionesco {\it et al.} for a presentation of these
experimental results and their discussion. 

We conclude this section by emphasising that perturbing fields, other than electric fields, are expected to 
lead to similar results. Studying non-linear responses seems to us a very promising route to follow in order to probe 
the glassy state in a new way. 
A particularly interesting case worth studying corresponds to non-linear mechanical responses of colloidal glassy liquids. In this case, the values of the perturbing field that affect the sample are of the order of Pa, 
thus much smaller than 
the ones affecting the measuring apparatus, which are of the order of GPa. Thus, within a very good approximation, the only non-linear output signal 
is from the sample itself. This is not the case for dielectric measurements, which are therefore very difficult since one has to be able to filter 
out the trivial non-linear part due to the amplifiers, etc. This was the main difficulty in the experiment reported in \protect\shortcite{chi3preprint}.

\subsection{Inhomogeneous dynamical susceptibilities}

We have seen in section \ref{inequality} that the variation of a dynamical correlator 
with respect to an external parameter (e.g., $\chi_T$) is a
way to obtain estimates of the number of dynamically correlated particles. As a natural
generalisation, one can study the variation of a local correlator, which
measures the dynamics around the position $\bx$, induced by a perturbation at certain other point $\bf z$.
By summing over all $\bz$ one obtains again a global dynamical response such as $\chi_T$, since this then corresponds 
to computing the variation with respect to a uniform shift of the external parameter.

This new, spatially dependent, dynamical response function is akin to $G_4$ and allows one to probe the spatial structure
of dynamic heterogeneity and to measure directly a
dynamical correlation length $\xi$. 
The physical reason is that {\it
spontaneous} dynamical fluctuations measured by the 4-point function
and {\it induced} dynamical fluctuations measured by
this new type of response function
are intimately related.  Accelerating or slowing down the dynamics at one
given point (by adding an external potential) must perturb the
dynamics over a length scale $\xi$ if the dynamics are indeed correlated
over this distance.

Let us define more precisely this new dynamical response. 
Consider the change in the local dynamical structure factor $ F(\bx,\by,t)$
due to an extra, spatially varying, external potential $U(\bz)$. Note that this observable
can always be decomposed in Fourier
modes: $ \frac{\delta F(\bx,\by,t)}{\delta
U(\bz)}\left.\right|_{U=0}=\int d\bk d\bq
e^{-i\bq\cdot(\bx-\by)+i\bk\cdot(\by-\bz)} \chi_{\bk}(\bq,t)$, where
$\chi_{\bk}(\bq,t) \propto \frac{\delta F(\bq,\bq+\bk,t)}{\delta
U(\bk)}\left.\right|_{U=0}$
is the response of the dynamical structure factor to a static external
perturbation in Fourier space.  For a perturbation localised at the
origin, $U(\bz)=U_0\delta(\bz)$, one finds $\delta
F(\bq,\by,t)=U_0\int d\bk e^{i\bk\cdot\by}\chi_{\bk}(\bq,t)$.  This susceptibility is also related to a 3-point density
correlation function in the absence of the perturbation. It is very important to note the very different role played by $\bq$ (the standard wavevector) and $\bk$ (the 
wavevector of the perturbation): only the latter is sensitive to dynamic correlations.\footnote{We note that compared to Refs.~\protect\shortcite{jcpI,IMCT}, 
the notations $\bq$ and $\bk$ have been inverted.}

The susceptibility 
$\chi_{\bk}(\bq,t)$ is interesting for experimental reasons, at least in colloids,
since it could be measured by using optical tweezer techniques. From a fundamental point
view, it provides a very useful way to characterise dynamic
heterogeneity since it is not affected by complications due to conservation 
laws and the type of dynamics, contrary to  $\chi_4$ and $G_4$. Therefore we expect that extracting spatial 
information and, especially, a precise estimate 
of $\xi$ should be cleaner by using 
 $\chi_{\bk}(\bq,t)$ (see the discussion in section \ref{caveats}). 
 
 Finally, another advantage of  $\chi_{\bk}(\bq,t)$ is that precise quantitative predictions
 have been obtained within MCT by analytical arguments \protect\shortcite{IMCT}, which were later confirmed by numerical analysis \protect\shortcite{Szamelchi4n} and complementary approaches \protect\shortcite{Szamelchi4a}. 
The critical behaviour of $\chi_{\bk}(\bq,t)$ approaching the MCT transition temperature is the following (we use the same 
MCT notation introduced previously):
\begin{itemize}
\item In the $\beta$-regime, i.e for times of the order of $\tau_\beta=\epsilon^{-1/2a}$, one finds
\begin{equation}
\chi_{\bk}(\bq,t)\propto \frac{1}{\sqrt{\epsilon}+\Gamma k^2} \, g_{\beta}\left(\frac{k^2}{\sqrt{\epsilon}},
t\epsilon^{1/2a}\right), 
\end{equation}
where the proportionality constant depends on $q$. The scaling function $g_{\beta}(k^2/\sqrt{\epsilon},t\epsilon^{1/2a})$ is regular for $k=0$, thus
implying that the $\bk=0$ value diverge as $1/\sqrt{\epsilon}$.  For 
large values of $u = t \epsilon^{1/2a}$ 
one finds that $g_{\beta}(k^2/\sqrt{\epsilon},t\epsilon^{1/2a})$ 
equals $\Xi(\Gamma
k^2/\sqrt{\epsilon})u^b$, with $\Xi$ a certain regular function.

\item In the $\alpha$-regime, i.e. for times of the order of $\tau_\alpha=\epsilon^{-1/2a-1/2b}$, one finds

\be
\label{latebeta} \chi_{\bk}(\bq,t) =\frac{\Xi(\Gamma
k^2/\sqrt{\epsilon})}{\sqrt{\epsilon} (\sqrt{\epsilon} +
\Gamma k^2)} \, g_{\alpha,q}\left(\frac{t}{\tau_\alpha}\right) 
\ee
with $\Xi$ is the same function defined previously. It has the properties: $\Xi(0) \neq 0$ and $\Xi(v
\gg 1) \sim 1/v$ such that $\chi_{\bk}$ behaves as $k^{-4}$ for large
$k\epsilon^{-1/4}$, independently of $\epsilon$. Also,
$g_{\alpha,q}(u \ll 1) \propto u^b $, as to match the $\beta$
regime, and $g_{\alpha}(u \gg 1,k) \to 0$.  
\end{itemize}

Note that the spatial scaling variable is $k^2/ \epsilon^{-1/2}$ in
both the $\alpha$ and the $\beta$ regimes. The physical consequence 
is that there exists
a unique diverging dynamic correlation length $\xi
\sim \sqrt{\Gamma} |\epsilon|^{-1/4}$ that rules the response of
the system to a space-dependent perturbation within MCT. The analysis of the
early $\beta$ regime where $t \ll |\epsilon|^{-1/2a}$ shows that
this length in fact first increases as $t^{a/2}$ and then saturates at
$\xi$. Interestingly, this suggests that while keeping a fixed
extension $\xi$, the (fractal) geometrical structures carrying the
dynamic correlations significantly ``fatten'' between
$\tau_\beta$ and
$\tau_\alpha$, where more compact structures are expected, as perhaps
suggested by the results of \protect\shortcite{Kob}. 

Up to now, there are no simulations or experiments measuring a spatial
dynamic response such as $\chi_{\bk}(\bq,t)$. Hopefully, these will be performed
in the future. As discussed previously, we do believe that this new observable 
is a simpler and more direct measure of dynamical correlations than $G_4$. 
Furthermore, quantitative results beyond scaling can be obtained within MCT. 
Thus one could consider comparing 
the MCT predictions for dynamical heterogeneities to numerical and experimental
result in a stringent way, as it has been done for the intermediate scattering function, see e.g. \protect\shortcite{KA}.

\subsection{Structure and dynamics: Is 
dynamic heterogeneity connected to the liquid structure?}

One of the most frequently asked questions in studies of dynamical
heterogeneity is whether the observed fluctuations might be 
structural in origin. Such questions have attracted sustained
interest. For example, in early numerical 
work on dynamic heterogeneity, immobile regions were discussed 
in connection with compositional fluctuations in fluid mixtures~\protect\shortcite{hurley}.
Thirteen years
later, some form of local crystalline order is 
invoked to account for slow domains in numerical
work~\protect\shortcite{tanaka-mrco}.

It should be noted that this chapter, and perhaps even this whole book about `dynamic
heterogeneity' would not exist in this form
if the question of the connection between 
structure and dynamics had been satisfactorily answered.
In that case, indeed, research would be dedicated to 
understanding the development of structural correlations
at low temperatures in supercooled liquids, and to developing 
tools to measure, quantify and analyse such static 
features. 

A key advance in connecting structural properties to dynamical heterogeneity
has been the development of the so-called `isoconfigurational
ensemble'~\protect\shortcite{harrowell1}.  
In this approach, one calculates a traditional ensemble average in two stages.
First, one averages the particles' velocities, keeping their
initial positions fixed.  (If the dynamics are stochastic, this step 
also contains an average over random noises.)  Averaging a local
dynamical observable 
such as $c(r;t,0)$ in this way, one arrives at an `isoconfigurational average'
$\langle c(r;t,0) \rangle_\mathrm{iso}$, which still depends on the position $r$
through the fixed initial particle positions.  This average is therefore
able to reveal the influence of the structure of the initial configuration on the dynamical
behavior at that point.  To return to a traditional ensemble average, one carries out
an average over the initial particle positions in a second step.

The right panel of Fig.~\ref{fig:dh_poster}
represents the spatial dependence of the 
isoconfigurationally averaged single-particle mobility in a two-dimensional mixture of soft 
disks~\protect\shortcite{harrowell1}. 
The fact that this image is not uniform 
demonstrates that part
of the dynamic heterogeneity has a structural origin. 
This raises two different questions. First, can one predict 
from structural measurements the pattern produced by the 
isoconfigurational average in Fig.~\ref{fig:dh_poster}?
Second, how much of the `real' dynamic heterogeneity is 
actually preserved by the isoconfigurational average and has thus
a genuine structural origin?

Harrowell and coworkers 
have provided detailed answers to the first 
question~\protect\shortcite{harrowell1,harrowell2,harrowell2b,harrowell3}. 
Statistical analysis of isoconfigurational 
ensembles has been very useful in assessing
the statistical significance of correlations between mobility
and the local energy, composition or free volume. 
They have recently made the point that strong correlations
exist between vibrational properties of the 
liquid and isoconfigurational mobilities~\protect\shortcite{harrowell4}.  
They have also 
made vivid the distinction between the existence of a statistical 
correlation between structural and dynamical fluctuations, and 
the much more demanding notion of a causal link between the two, 
that is, of a correlation that is strong enough that 
prediction of the mobility can be made based on a given structural 
information~\protect\shortcite{harrowell2,harrowell2b}. 
These two notions are very often confused in 
the dynamic heterogeneity literature. 

The example of enthalpy fluctuations
is useful in this respect. The fact that the four-point
susceptibility $\chi_4(t)$ can be quantitatively
estimated with good accuracy from a three-point susceptibility such as 
$\chi_T(t) \propto \langle \delta H(0) \delta C(t,0) \rangle$
provides evidence of a strong correlation between enthalpy
and dynamic fluctuations. However, enthalpy fluctuations 
are not good predictors for dynamic heterogeneity, presumably because
they contain short-ranged and short-lived 
fluctuations that do not correlate well with slow dynamics. 
Indeed, suitably filtered enthalpy fluctuations
correlate very strongly  with dynamic heterogeneity~\protect\shortcite{poole}.

We finally return to the second question: 
is dynamic heterogeneity truly captured by isoconfigurational 
averages, and thus does it fully originate from the structure?
The response is more subtle than expected as it depends on which
observable, and more precisely on which lengthscale, 
it is analysed. We mentioned above that dynamic heterogeneity
primarily revealed itself through the intermittent single particle
dynamics (Fig.~\ref{msd2}) leading to broad distributions
of single particle displacements with broad tails. These features almost
completely disappear after the isoconfigurational average
is performed~\protect\shortcite{robludo}. In other words, the distinction 
between mobile and immobile particles is mostly dynamical in nature,
suggesting that the quest for a connection between
the static and dynamic properties of glass-formers at 
the particle level is in vain. 

Nevertheless,
 mobility fluctuations
do display interesting spatial correlations, as illustrated in Fig.~\ref{fig:dh_poster}.
This suggests that
the distinction between fast and slow domains remains 
consistent in the isoconfigurational ensemble. This observation 
can be quantified by measuring a `restricted' four-point function
\be
\chi_4^{\rm iso}(t) = N  \bigg\langle  
\langle C(t,0)^2 \rangle_{\rm iso} - \langle C(t,0) 
\rangle^2_{\rm iso} \bigg\rangle_{\rm initial \,\, cond.}.
\ee 
While $\chi_4(t)$ measures the total strength 
of dynamic heterogeneity, $\chi_4^{\rm iso}(t)$ makes use of the 
isoconfigurational ensemble and first records the strength of 
dynamic heterogeneity at fixed initial conditions, the average
over initial conditions being performed afterwards. In the
case where isoconfigurational mobility 
(and thus the image in Fig.~\ref{fig:dh_poster}) is uniform,
one has $\chi_4^{\rm iso}(t) = \chi_4(t)$, since the average over initial 
conditions is trivial in this case. More generally, a large
contribution of $\chi_4^{\rm iso}(t)$ to $\chi_4(t)$ indicates 
that the dynamic fluctuations captured by $\chi_4(t)$ are inherently dynamical 
in origin and do not originate in the liquid structure. Numerical measurements in molecular
dynamics simulations indicate that the opposite is true and 
$\chi_4^{\rm iso}(t)$ contributes less and less to $\chi_4(t)$ as 
temperature decreases~\protect\shortcite{robludo}. This suggests that the search 
of a causal link between structure and mobility does make sense, at least on 
 large length scales. Interestingly, the vibrational 
properties investigated in Ref.~\protect\shortcite{harrowell4} as relevant 
structural indicators of dynamic heterogeneity are 
a suitable candidate, since the vibrational spectrum is a collective 
property. 

\subsection{Point to set correlations: 
Emergence of amorphous long range order?}
\label{pointtoset}

As discussed in the previous section, it is quite natural to ask what structural features (if any) might be responsible
for the growth of dynamical correlations. One possibility is that actually there exists a static growing length that 
drives the increase of dynamical correlations. As we discussed in the introduction, simple static correlations are rather 
featureless when approaching the glass transition. However a new length called the ``point to set'' length was recently introduced 
 \protect\shortcite{bbpsi,MM}. 
It is naturally devised to probe the growth of static amorphous long range order \protect\shortcite{bbpsi,MM} and has been
shown to grow close to the glass transition \protect\shortcite{NP}.
This is reviewed in the chapter by Semerjian and Franz.

The basic idea is to measure how much boundary conditions affect the behaviour of the system, far away from the boundaries themselves. 
This is the usual
way to test for the emergence of long range order in statistical mechanics. However, for standard phase transitions, 
the appropriate boundary conditions are known
from the outset. For example, in the case of ferromagnetic transitions, one 
can fix the boundary spins mostly in the up direction and check whether this leads to a positive magnetisation for spins in the bulk. 
The difficulty in the case of glasses, for which one would like to test the presence of long range amorphous order, is that the 
boundary conditions one has to use look just as random as the amorphous configuration one wants to select. 
The way out is to let the system itself choose the boundary conditions: the procedure is to take an equilibrated configuration $\alpha$, 
freeze all particles outside a cavity of radius ${\mathcal R}$ and then recompute the thermodynamics for the particles
inside the cavity, that now are subjected to a typical equilibrium boundary condition. One can then study 
a suitably defined average overlap $q({\mathcal R})$ between the new thermalized configurations at the centre of the cavity and the reference state $\alpha$, as a function of ${\mathcal R}$. 
The quantity $q({\mathcal R})$ is  called a ``point-to-set'' correlation \protect\shortcite{MM}. The characteristic length scale over which $q({\mathcal R})$ drops to zero is 
called the point to set length. The increase of this length is a clear signal that the system is developing long range static order, and in the case of glasses, 
amorphous long range order. This point-to-set length is precisely the size of the glassites within the RFOT theory of glasses, see section \ref{rfot}. 

This topic is discussed in detail by
Franz and Semerjian, to which we refer for a presentation 
of the general theoretical and numerical results.
Here we simply mention that this point to set length 
has been shown to grow in numerical simulations of 
supercooled liquids \protect\shortcite{NP}. 
Furthermore, it was proved that it must diverge whenever the 
relaxation time does so~\protect\shortcite{SM}. 

Therefore, an important open question is whether the correlation length picked up by the dynamical correlators discussed above 
are actually just a consequence of hidden static correlations or if they are instead 
quite unrelated to them. In the first
case, the study of dynamical correlation will still be very valuable because it provides an easier way to probe static correlations. 
Whatever is the correct answer, it would lead to a substantial progress in our understanding of the glass transition and would help us in pruning down the correct theory. 
We do hope that numerical and experimental studies will be devoted
to this important problem in the future.      

\subsection{Large deviations and space-time thermodynamics}

It is clear from its definition in Eq.~(\ref{equ:g4def}) that
the correlation $G_4(r,t)$ is the covariance of the mobility
$c(r;0,t)$ at two nearby spatial points.  Similarly,
the three-point functions of the previous section are covariances
of $c(r;0,t)$ with the local energy, enthalpy or free volume.
Of course, not all information about the mobility is contained
in such covariances: one might consider higher moments
of these functions or indeed the joint distribution of mobilities
at all points.
However, the inherent difficulty of characterising the distribution
of an entire mobility field requires physical intuition in 
choosing which observable to measure.

\begin{figure}[t]
 \includegraphics[width=4.1cm,clip]{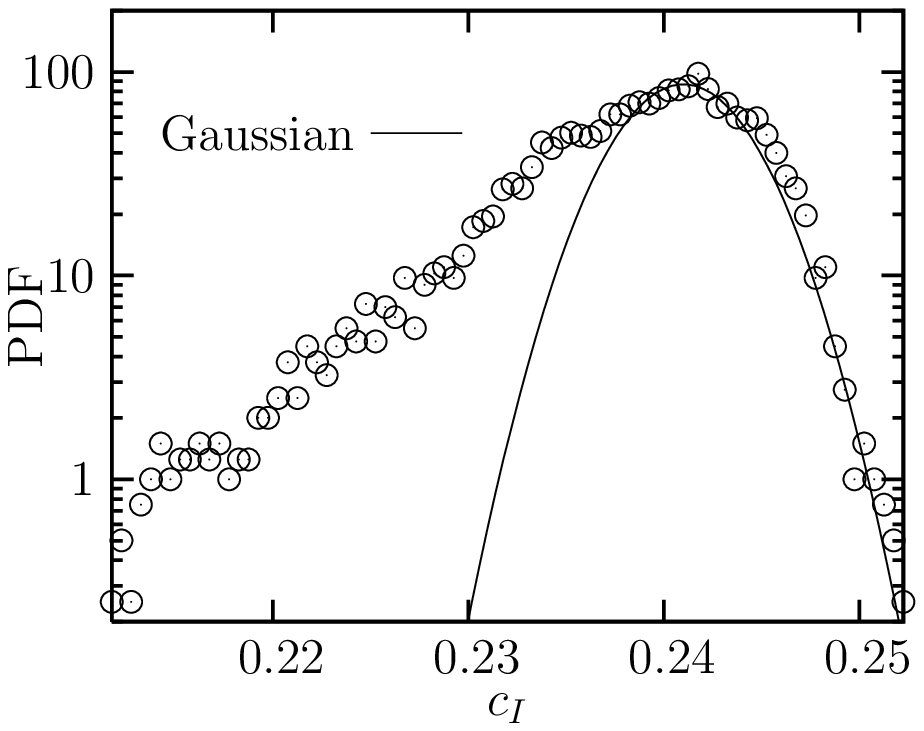}
 \includegraphics[width=3.5cm,clip]{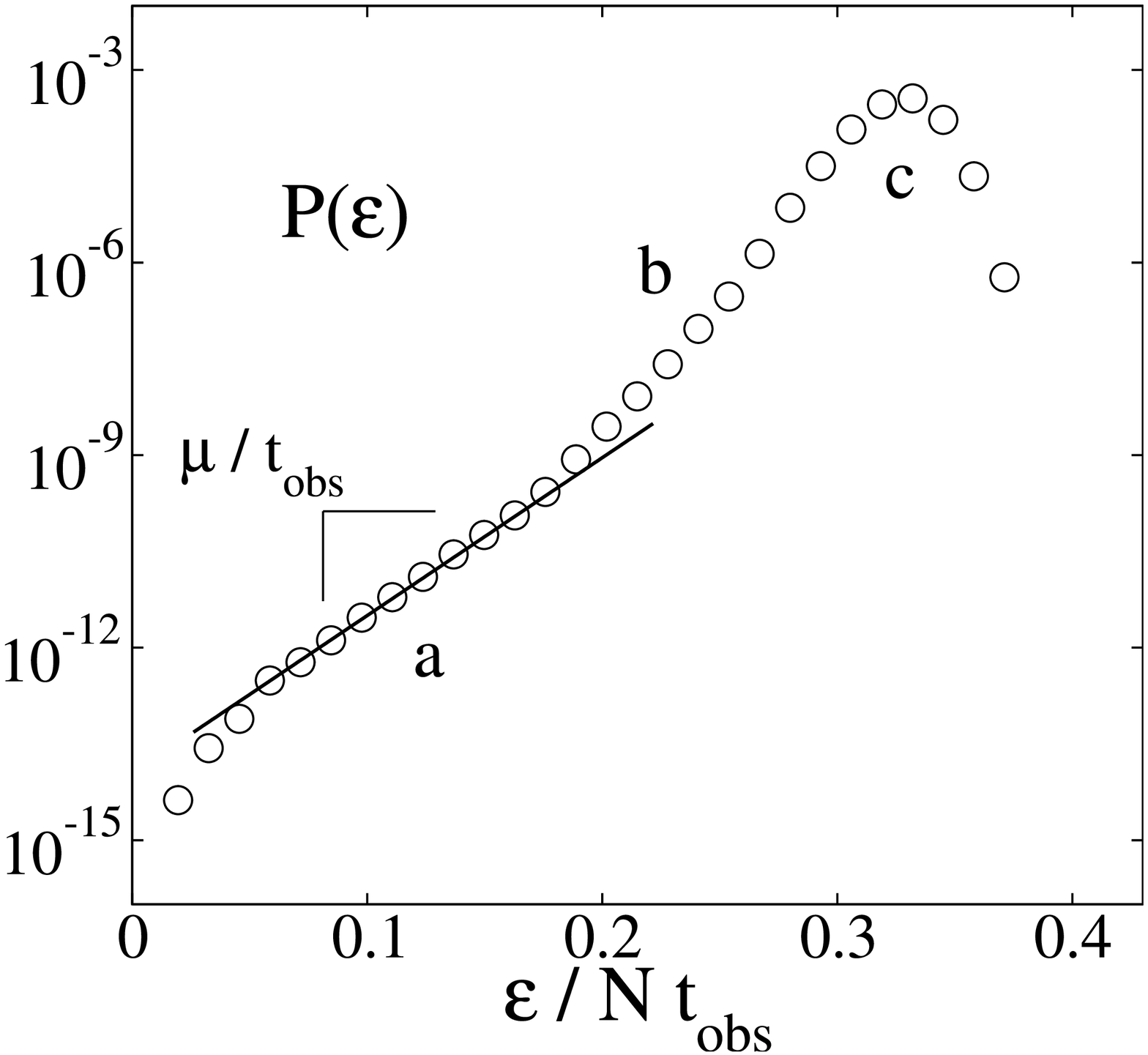}
 \includegraphics[width=3.9cm,clip]{fig10c.ps}
\caption{Distributions of dynamical observables.
(Left) Distribution of the correlation function in a coarsening
foam~\protect\shortcite{luca-dist}.
The skewed distribution arises from rare trajectories
with more motion than average.
(Centre) The distribution of the `dynamical action'
$\varepsilon$ in a kinetically constrained model~\protect\shortcite{merolle}.
The action $\varepsilon$ measures the amount of motion in a trajectory
in the same spirit as the activity $K$ defined in Eq.~(\ref{equ:def_K}). 
The distribution is left-skewed and non-convex, 
indicating a population of trajectories with low mobility. 
(Right) Distribution of the activity $K$ in a model of Lennard-Jones
particles, with a biasing field $s$ in place, as discussed in the text.
Two peaks are evident when the $t_\mathrm{obs}$ is large, revealing
the presence of two distinct dynamical phases~\protect\shortcite{hedges-science}.
} \label{fig:merolle-etc}
\end{figure}

A natural first choice for such a scheme is to consider
the fluctuations of the spatially averaged correlation
function $C(t,0)$, beyond the Gaussian level.  Such
measurements are possible in experiments such as those
of Duri \emph{et al.}~\protect\shortcite{luca-dist} (see Fig.~\ref{fig:merolle-etc})
and in computer simulations of a variety of models~\protect\shortcite{chamon-reichman}. 
They have typically been considered in out-of-equilibrium 
situations but
this is not essential (see also below).  Typically,
such distributions are skewed and non-Gaussian, and it
is natural to connect the asymmetry of the distribution
with dynamical heterogeneity.  For example, even on 
time scales $t$ much greater than the structural relaxation
time, there is a substantial probability that regions
of the system have persisted in an immobile state for all
times between 0 and $t$.  This enhances the probability
of observing a larger than average value for $C(t,0)$.
The opposite behaviour may
occur on short times: while typical regions have not relaxed,
co-operative motion in some regions enhances the probability
of observing a smaller than average of $C(t,0)$.

\newcommand{\tobs}{t_\mathrm{obs}}

In making this connection,
it seems reasonable that regions where $C(t,0)$
is large possess rather stable structure at the molecular level,
while regions where $C(t,0)$ is small correspond to relatively
unstable local structure.  This fact has recently been exploited 
in computational studies that probe trajectories where relaxation
is much slower than average.  
To identify such trajectories, it is useful to define a measure
of dynamical activity, for systems of $N$ particles evolving
over an observation time $t_\mathrm{obs}$.  For example, one may
take~\protect\shortcite{hedges-science}
\be
K(N,t_\mathrm{obs}) = \sum_{i=1}^N \sum_{j=1}^{\tobs/\Delta t} 
|\bm{r}_i(t_j) - \bm{r}_i(t_j-\Delta t)|^2,
\label{equ:def_K}
\ee
where $\bm{r}_i(t)$ is the position of particle $i$ at time
$t$ and the $t_j=j\Delta t$ are equally spaced times.
For large $N$ and $t_\mathrm{obs}$, the distribution of $K$
becomes sharply peaked about its average,
$\langle K \rangle$.  

In general,
for large $N$ and $t_\mathrm{obs}$, one expects a the equilibrium
distribution of $K$ to have the form 
\be
P(K) \simeq \exp[-N t_\mathrm{obs} f(K/Nt_\mathrm{obs})]
\ee
where the function $f(k)$ resembles a free energy density:
it gives the probability of observing a substantial
deviation between the measured $K$ and its average 
$\langle K \rangle$.  
(The variance of $K$ was also considered 
in Ref.~\protect\shortcite{merolle}.
In general, this quantity contains different
information to four-point
functions such as $\chi_4(k,t)$ although $\chi(\tobs)$ and $\chi_4(k,t)$
may sometimes be related through scaling arguments\footnote{
In the notation of Eq.~(\ref{equ:g4def}), one may write $K=\sum_{i} \sum_j c_i(t_j,t_j+\Delta t)$
so that the variance of $K$ contains terms like $g_{ijmn} = \langle c_i(t_j,t_j+\Delta t) c_m(t_n,t_n+\Delta t)\rangle$.  
Assuming that $g_{ijmn}$ depends on scaling variables such as $|r_i-r_j|/\xi_4$ and $(t_j-t_n)/\tau_\alpha$,
one may connect the variance of $K$ to the dynamical length scale $\xi_4$ and time scale $\tau_\alpha$.  
Such connections are
are analogous to the relation (\ref{directlink}) 
between $\chi_4(t)$ and $\xi_4(t)$, but there is considerable freedom in the scaling ansatz for $g_{ijmn}$,
which seems to prevent a more 
direct connection between the variance of $K$ and $\chi_4(t)$.
}.)
In some kinetically constrained models~\protect\shortcite{merolle}, the distribution $P(K)$
has a characteristic shape, skewed towards small activity,
with an apparently exponential tail, as shown in 
the central panel of Fig.~\ref{fig:merolle-etc}.
Further, on estimating $f(k)$ from this plot, there is a 
range of $K$ over which $f(k)$ is non-convex (that is,
$f''(K)<0$).  The behaviour of $f(k)$ away from
its minimum describes the properties
of rare trajectories in the system and their relevance for
the liquid behaviour is not clear \emph{a priori}.  However,
the key motivation of this
study was to formulate a thermodynamic approach to
the statistical properties of trajectories of glassy 
systems~\protect\shortcite{GCprl}.
Within such a framework, non-convexity of $f(K)$ has
a direct interpretation as a `dynamical phase transition' in the system.

This interpretation is most easily seen by taking the Legendre transform
of $f(K)$ to obtain a new `dynamical free energy'
\be
\psi(s) = -\min_k [ sk + f(k) ],
\ee
which describes the response of the system to a field
$s$ that biases the system towards trajectories with
small (or large) activity $K$.  In particular, the effect
of the bias is to change the average of $K$ from
$\langle K\rangle$ to $K(s) = 
-N\tobs \frac{\mathrm{d}}{\mathrm{d}s}\psi(s)$.
Then, a non-convex form for $f(K)$ results in a 
jump singularity of $K(s)$ for a specific biasing field $s=s^*$.
While the field
$s$ has no simple physical interpretation, one may view it
as a mathematically convenient trick for sampling the distribution $P(K)$.

Turning to the results of this formalism, the key point is that glassy
systems may exhibit singular responses to the field $s$, leading to
`ideal glass' states that are characterised by values of
$K$ that is much smaller than its equilibrium average
$\langle K \rangle$.  The existence of 
these phase transitions has been proven in simple
models~\protect\shortcite{garrahan-fred,jack-rom} 
and numerical results for Lennard-Jones model
liquids are also consistent with the existence
of such a transition~\protect\shortcite{hedges-science}.  In particular, if the field $s$
is chosen to lie at the putative phase transition point, then one may 
construct the distribution of $K$ in the presence of the field $s$,
$P_s(K) \propto P(K) {\rm e}^{-sK}$.
In the presence of a phase transition, $P_s(K)$ has two peaks, which
correspond to distinct active (liquid) and inactive (glass) states.
An example is shown in the right panel of Fig.~\ref{fig:merolle-etc}.

To summarise then, dynamical heterogeneity is concerned with 
distributions of dynamical quantities, most often
through their means and covariances.  However, the tails
of these distributions can reveal information
about possible new phases in the system, whose structure
is very stable and whose relaxation times are very long.  
This leads to the hypothesis that the nature of the
dynamically heterogeneous fluid state should be interpreted
in terms of coexistence between and active liquid and 
inactive `ideal glass' states.

\section{Open problems and conclusions}

The aim of this chapter was to review the recent progress in the 
quantitative analysis of dynamical heterogeneities. We showed that 
the introduction of four-point correlation functions played an important 
role, both conceptually and operationally, by providing a precise quantitative measure
to characterise dynamical heterogeneities. These four-point correlations have
now been measured or estimated in numerical simulations of schematic and realistic
models of glass formers, and experimentally on molecular glasses, colloids and
granular assemblies close to jamming. They have also been investigated theoretically 
within simplified models or within the mode-coupling approximation, and have indeed been shown
to be critical as the glass transition is approached. These four-point correlations are the
natural counterpart, for glass or spin-glass transitions, of the standard two-point correlations
that diverge close to a usual second-order phase transition.  These can be considered to be 
breakthroughs that have significantly improved our understanding of the microscopic mechanisms 
leading to glass formation, and that have already spilled over to many different scientific communities.

However, it soon became clear that four-point correlation functions are not a panacea. It was for example
not anticipated that these functions would be delicately 
sensitive to details such as the choice of 
statistical ensemble or the microscopic dynamics. Second, although these four-point objects give valuable 
information, they are not powerful enough to answer more precise questions about the geometry of the structures that
carry dynamical heterogeneities, or about the nature of the relaxation events (continuous vs. activated). 
Along the same line of thought, the relation between the dynamical correlation length extracted from these 
four-point points and the intuitive (but not so clearly defined) notion of cooperative relaxation is at this stage 
quite elusive. How many different `dynamical' length scales does one expect in general?

We have seen that the study of three-point response functions and non-linear susceptibilities allows one to bypass 
some of the difficulties inherent to four-point functions. We note that the experimental and numerical situation on
that front is much less developed, and should be encouraged. The very recent measurement of the non-linear dielectric 
properties of glycerol~\protect\shortcite{chi3preprint} is a remarkable exception. 

Higher-order correlation functions might also contain interesting 
quantitative information about dynamical heterogeneities, but
this subject is at this stage totally unexplored. It was recently 
suggested \protect\shortcite{levy} that six-point functions might provide
a way to measure intermittent dynamics and identify activated events. 
Skewness (or kurtosis) might indeed detect that the dynamics is 
intermittent,  as one expects if
`activated' events dominate, with a few rare events 
decorrelating the system completely,
while most events decorrelate only weakly.
More work in that direction would certainly be worthwhile.

Finally, we have not touched upon the problem of out-of-equilibrium dynamical heterogeneities, in particular in the 
aging regime. This is clearly a very interesting topic, for which experimental efforts are underallocated, 
although results in this regime might be able to 
discriminate between theories. We refer to 
\protect\shortcite{parisi,cugliandolo-chamon,castillo,lee1,lee2,djamel,weeksaging} 
and the chapter on aging in this book for 
interesting lines of research on this issue in spin-glasses.

\bibliographystyle{OUPnamed_notitle}

\bibliography{refs_chap2}

\end{document}